\documentclass[useAMS,usenatbib,aas_macros]{mn2e}
\usepackage{graphicx}
\usepackage{epstopdf}
\usepackage{bm}
\usepackage{times}
\usepackage{amsmath,amssymb}
\usepackage{slashed}
\usepackage{color}
\usepackage{natbib}
\usepackage{color}
\usepackage{hyperref}
\usepackage[]{appendix}
\usepackage{multirow}
\usepackage{tabularx}

\def\apj{\ref@jnl{ApJ}}                 
\def\apjs{\ref@jnl{ApJS}}               
\def\mnras{\ref@jnl{MNRAS}}             
\def\aap{\ref@jnl{A\&A}}                

\begin{document}

\title[A 3.5 keV line in the Galactic Center and a Critical Look at the Origin of the Line Across Astronomical Targets]{Discovery of a 3.5 keV line in the Galactic Center and a Critical Look at the Origin of the Line Across Astronomical Targets}

\author[T. Jeltema and S. Profumo]{Tesla Jeltema$^{1}$\thanks{tesla@ucsc.edu} and Stefano Profumo$^{1}$\thanks{profumo@ucsc.edu}\\
$^{1}$Department of Physics and Santa Cruz Institute for Particle Physics
University of California, Santa Cruz, CA 95064, USA
}


\maketitle

\begin{abstract}
We examine the claimed excess X-ray line emission near 3.5 keV including both a new analysis of {\em XMM-Newton} observations of the Milky Way center and a reanalysis of the data on M~31 and clusters. In no case do we find conclusive evidence for an excess.  In the case of the Galactic center we show that known plasma lines, including in particular K XVIII lines at 3.48 and 3.52 keV, provide a satisfactory fit to the {\em XMM} data. We estimate the expected flux of the K XVIII lines and find that the measured line flux falls squarely within the predicted range based on the brightness of other well-measured lines in the energy range of interest and on detailed multi-temperature plasma models. We then re-assess the evidence for excess emission from clusters of galaxies, allowing for systematic uncertainty in the expected flux from known plasma lines and additional uncertainty due to potential variation in the abundances of different elements. We find that no conclusive excess line emission can be advocated when considering systematic uncertainties in Perseus or in other clusters. We also re-analyze the {\em XMM} data for M~31 and find no statistically significant line emission near 3.5 keV to a level greater than one sigma.  Finally, we analyze the Tycho supernova remnant, which shows similar plasma features to the sources above, but does not host any significant dark matter.  We detect a 3.55 keV line from Tycho, which points to possible systematic effects in the flux determination of weak lines, or to relative elemental abundances vastly different from theoretical expectations.
\end{abstract}

\begin{keywords}
X-rays: galaxies; X-rays: galaxies: clusters; X-rays: ISM; line: identification; (cosmology:) dark matter
\end{keywords}   

\section{Introduction}
The particle nature of the dark matter, comprising most of the gravitationally bound structures in the universe, is unknown. A far-ranging experimental and observational program is in place to search for non-gravitational signals that could point to a given class of particle dark matter candidates. While weakly interacting massive particles have attracted much attention, other particle candidates remain theoretically robust and observationally viable. Among such candidates, ``sterile'' neutrinos offer the appealing possibility of tying the dark matter problem to the issue of generating a mass for the Standard Model ``active'' neutrinos, provide an interesting warm dark matter candidate, and can be potentially associated with a mechanism to explain the baryon-antibaryon asymmetry in the universe \citep[see ][for a recent review]{Boyarsky:2009ix}.

Sterile neutrinos can mix with active neutrinos and decay, on timescales much longer than the age of the Universe, to the two-body final state given by an active neutrino and a photon. The details of such process depend on the particular extension to the Standard Model that accommodates the sterile neutrino(s) \citep[see e.g. ][]{Pal:1981rm}, but the lifetime is set by a model-independent combination of the sterile-active neutrino mixing angle $\theta$ and of the sterile neutrino mass $m_s$ of the form
\begin{equation}\label{eq:sterilenu}
\tau\simeq7.2\times 10^{29}\ {\rm sec}\left(\frac{10^{-4}}{\sin(2\theta)}\right)^2\left(\frac{1\ {\rm keV}}{m_s}\right)^5.
\end{equation}
Such a decay mode produces an almost monochromatic photon signal at an energy approximately equal to half the sterile neutrino mass. Cosmological production mechanisms and constraints from phase-space density restrict the relevant range for the sterile neutrino mass to, roughly, 0.5 -- 100 keV \citep{Boyarsky:2009ix}. As a result, the expected line from sterile neutrino two-body decay falls in the X-ray range.

Earlier this year, \citet{Bulbul:2014sua} claimed the existence of an unidentified emission line at $E=(3.55-3.57)\pm0.03$ keV from stacked  {\em XMM-Newton} observations of 73 galaxy clusters with redshift ranging between 0.01 and 0.35. The line is observed with statistical significance greater than 3$\sigma$  in three separate subsamples: (i) the single Perseus cluster; (ii) the combined data for the Coma, Centaurus and Ophiuchus clusters; (iii) all other 69 clusters. Chandra observations of Perseus indicate a line feature compatible with the {\em XMM} results; The line was not, however, observed in the Virgo cluster with Chandra data. \citet{Bulbul:2014sua} explored possible contaminations from metal lines, notably from K and Ar, which they claim would require typical emissivities factors of 10-30 larger than expected from other bright lines.

Shortly after the analysis of \citet{Bulbul:2014sua}, a 3.5 keV line was reported from {\em XMM-Newton} observations of both the Perseus cluster and the Andromeda galaxy, while not being observed in ``blank sky'' observations \citep{Boyarsky:2014jta}. The line intensity is compatible with a sterile neutrino with a mass of $7.06\pm0.05$ keV, and a mixing angle $\sin^2(2\theta)=(2.2 -20)\times10^{-11}$ \citep{Boyarsky:2014jta}, consistent with the results of \citet{Bulbul:2014sua}, which quote a mass of about 7.1 keV and a mixing angle $\sin^2(2\theta)\sim7\times10^{-11}$.

Significant excitement from the model-building community followed the observations described above. Several studies focused on the possibility of decaying sterile neutrinos, and especially on the question of the genesis of the right abundance of such particles in the early universe and on the embedding of models in extensions to the Standard Model of particle physics \citep{Ishida:2014dlp, Abazajian:2014gza, Baek:2014qwa, Tsuyuki:2014aia, Allahverdi:2014dqa, Okada:2014vla, Modak:2014vva, Cline:2014eaa, Rosner:2014ch, Robinson:2014bma, Abada:2014zra}. Other studies considered alternative possibilities, including exciting dark matter \citep{Finkbeiner:2014sja, Okada:2014zea}, axion and axion-like dark matter \citep{Higaki:2014zua, Jaeckel:2014qea, Lee:2014xua, Cicoli:2014bfa, Ringwald:2014vqa}, axino dark matter \citep{Kong:2014gea, Choi:2014tva, Dias:2014osa, Liew:2014gia, Conlon:2014xsa}, gravitino dark matter \citep{Bomark:2014yja, Demidov:2014hka}, decaying moduli \citep{Nakayama:2014ova}, light vector bosons \citep{Shuve:2014doa}, R-parity violating \citep{Kolda:2014ppa} or R-parity conserving \citep{Dutta:2014saa, Baer:2014eja} decays of sparticles in supersymmetry, Majoron dark matter \citep{Queiroz:2014yna}, magnetic dark matter \citep{Lee:2014koa}, dark transition electric dipoles \citep{Geng:2014zqa}, or effective theory constructions \citep{Krall:2014dba} \citep[see also][for a general model-building discussion]{Frandsen:2014lfa}. The possibility of dark matter pair-annihilation was also entertained in \citet{Dudas:2014ixa} and \citet{Baek:2014poa}.

More recently, \citet{Riemer-Sorensen:2014yda} used Chandra observations of the Milky Way center to probe the possibility that the 3.5 keV line originates from dark matter decay. No evidence was found for excess X-ray line emission in the energy of interest when including lines at the energies of known plasma lines. When including known emission lines, \citet{Riemer-Sorensen:2014yda} rules out at the 95\% confidence level a dark matter decay scenario as the origin of the line signal reported by \citet{Bulbul:2014sua} and \citet{Boyarsky:2014jta}.  However, as this paper does not quote measured fluxes for the plasma emission lines near 3.5 keV, it is unclear if these are consistent with what is expected or could instead mask a dark matter signal.

While the results of \citet{Riemer-Sorensen:2014yda}, with the above caveat, put significant pressure on a dark matter decaying interpretation of the X-ray line observed from Galaxy clusters, \citet{Conlon:2014wna} point out that the magnetic field structure of the Milky Way is such that an axion-like particle conversion to a monochromatic X-ray photon in the presence of magnetic fields would still be a viable option. \cite{Conlon:2014xsa}, in fact, found that for such a case the line intensity predicted, for instance, for M31 is two orders of magnitude larger than for the Milky Way. 

Four key subsequent observational studies cast serious doubt on an exotic origin for the line signal of \citet{Bulbul:2014sua} and \citet{Boyarsky:2014jta}. \cite{Malyshev:2014xqa} considered stacked observations of dwarf galaxies, while \cite{Anderson:2014tza} analyzed Chandra and {\em XMM-Newton} observations of two samples of galaxies and groups of galaxies; such a sample possesses the important feature that no significant plasma emission should exist at 3.5 keV. Neither one of those studies found any evidence for a 3.5 keV line signal, robustly ruling out a dark matter decay interpretation of the 3.5 keV line observations reported by \cite{Bulbul:2014sua} and  \cite{Boyarsky:2014jta}. Subsequently, \cite{Urban:2014yda} utilized deep deep Suzaku observations of Perseus, Coma, Virgo and Ophiuchus; while a 3.5 keV line was significantly detected in the case of Perseus, with an intensity  potentially explained by elemental lines, no signal was found from the  other three clusters, ruling out a dark matter interpretation for the 3.5 keV line observed from Perseus, when appropriately rescaled for the other three clusters. Finally, with Carlson, we recently showed that the morphology of the Galactic Center and of the Perseus 3.5 keV signal is entirely incompatible with a dark matter decay origin, while being consistent with an origin rooted in plasma emission in both cases, and we derived the most stringent constraints to date on the sterile neutrino parameter space for masses in the vicinity of 7 keV \citep{Carlson:2014lla}.

In the present study we utilize {\em XMM-Newton} observations of the Milky Way center and of M31 with the purpose of testing a ``new physics'' origin for the 3.5 keV line reported by \cite{Bulbul:2014sua} and \cite{Boyarsky:2014jta}. We illustrate that known plasma emission lines, with reasonable emissivities, provide a satisfactory fit to the Galactic center X-ray spectrum in the energy range of interest (sec.~\ref{sec:plasma}) and we outline the implications for a dark matter interpretation of the 3.5 keV line (sec.~\ref{sec:dm}); we then show that accounting for systematic uncertainties in the plasma emission line brightness in clusters of galaxies from both relative elemental abundances and plasma multi-temperature models explains the observed 3.5 keV feature (sec.~\ref{sec:clusters}); finally, we show that there is no statistical evidence, to a level greater than one sigma, for the existence of a 3.5 keV line from M~31 in a reanalysis of the {\em XMM} data. (sec.~\ref{sec:m31}). We discuss possible systematic effects in sec.~\ref{sec:sys}, and summarize our findings and conclude in sec.~\ref{sec:conclusions}.

\section{Searching for Dark Matter Decay from the Milky Way center}

As a starting point to our analysis, we focus on {\em XMM} observations of the Galactic center. The center of the Milky Way is an obvious target to search for non-gravitational signals from particle dark matter, such as photons from dark matter pair annihilation or decay. While astrophysical background emission, including significant flaring activity, is present in the region, simple estimates of the integrated line of sight density of dark matter within angular regions of the size of the field of view of X-ray instruments indicate that the Galactic center is likely the brightest direction in the sky for dark matter decay into photons. Here, we improve on a recent analysis that utilized {\em Chandra} X-ray observations of the Galactic center \cite{Riemer-Sorensen:2014yda} by employing archival public {\em XMM} observations, with a larger field-of-view, a higher effective area, and with a total cleaned exposure time a factor almost 2.5 greater than in the analysis of \cite{Riemer-Sorensen:2014yda}.  We also explore in detail the expected contribution of known plasma lines and whether these can explain the observed emission.

\subsection{{\em XMM} Data Analysis} \label{subsec:xmm}

\begin{table*}
\centering
\begin{tabular}{cccc} \hline
obsID  &      MOS1 (ksec) &  MOS2 (ksec) &  PN (ksec)  \\
\hline
0111350101  &  41.5 &  41.2	& 34.6  \\
0202670701    &  79.9	&  83.4	& 53.6\\
0554750401    &  31.7	&  33.3	& 26.1\\
0604301001    &  39.2  & 41.2  & 20.9\\
0674601101    &  10.6	&  10.9&	 9.4\\
0202670801    &  93.3 &	  97.0 &	 62.5 \\
0554750501    & 40.6	&  40.6	&  31.6\\
0658600101   &  47.1	&  47.6 &	 39.8\\
0554750601   &  36.3	&  36.3 &	 24.8\\
0658600201    &  39.7	&  42.7 &	 32.6\\
0604300601     &  30.7	&  32.2 &	 21.0\\
0112972101   &   21.3	&  21.3 &	 16.9\\
0604300701    &  36.6	&  42.1 &	 19.9\\
0604300801    &  34.6	&  34.6 &	 27.6\\
0674600801    &  18.0	&  18.4 &	 13.8\\
0604300901    &  20.9	&  22.1 &	 13.5\\
0674601001    &  20.6	&  21.1 &	 15.5\\
0202670601    &  32.2	&  34.8 &	 22.4\\
\hline
\end{tabular}
\caption{{\em XMM} {\tt obsID} numbers and flare-filtered exposure times for the three EPIC cameras for the Galactic center observations used in our analysis.}
\label{obslog}
\end{table*}

In this section, we describe the observations we employed in our analysis, the data reduction and selection procedure. 
We initially considered all {\em XMM} observations pointed within 4' of the Galactic center and with exposure times of at least 10 ksecs.  The EPIC MOS and PN data for all observations were reduced with the {\em XMM} SAS (version 13.5.0) software\footnote{http://xmm.esac.esa.int/sas/} using standard reduction techniques.  The level 1 event files were reprocessed with the {\tt emchain} and {\tt epchain} tasks.  Flare filtering, point source detection, and spectral extraction were carried out  using the {\em XMM} {\tt ESAS} package \citep{esas,esas2}.  The lightcurve filtering tasks {\tt mos-filter} and {\tt pn-filter} within {\tt ESAS} are designed to eliminate periods with elevated particle background.  Filtering was accomplished by creating a histogram of the count rate in 60 second bins; a Gaussian was fit to the histogram, and periods when the count rate deviated by more the 1.5$\sigma$ from the Gaussian peak were removed.

In the case of the Galactic center, Sgr A$^{\ast}$ is known to be highly variable, and, in addition to actual particle background flares, flaring activity of Sgr A$^{\ast}$ can also significantly change the observed count rate.  In principle, flaring of Sgr A$^{\ast}$ should have no effect on our analysis, since it would not affect the flux of a dark-matter-induced line.  However, these flares do significantly increase the background to line detection, as well as changing the spectral shape of Sgr A$^{\ast}$ \citep[e.g.][]{2012ApJ...759...95N}, so we removed from our analysis observations during time periods with strong Sgr A$^{\ast}$ flares (October 3, 2002; April 2-5, 2007) as well as additional observations which were found to be very highly contaminated by variability and/or particle background flares.  The observations utilized in our analysis and the remaining good exposure time after flare filtering for each instrument are catalogued in Table~\ref{obslog}.  The total flare-filtered exposure times are 675 ks, 700 ks, and 487 ks for the MOS1, MOS2, and PN detectors, respectively. Collectively, this exposure time is approximately 2.5 times larger than the {\em Chandra} observations used in the analysis of \cite{Riemer-Sorensen:2014yda}.

Bright point sources were detected and masked using the {\tt ESAS} task {\tt cheese} run on broad band images (0.4-7.2 keV), including a $\sim20"$ region around Sgr A$^{\ast}$; cheese also masks low exposure regions of the detector like chip gaps and bad columns.  We then extracted spectra from the full field-of-view for each detector, excluding CCD 6 on MOS1 which suffered micrometeoroid damage in March 2005, and corresponding redistribution matrix files (RMFs) and ancillary response files (ARFs) using the {\tt ESAS} tasks {\tt mos-spectra} and {\tt pn-spectra}.  No background spectra were created, as we choose to model the background rather than subtract it off.  The individual spectra were then summed using the {\tt mathpha} tool from the {\tt FTOOLS} package \citep{ftools}.  Combined RMF and ARF files are generated using the {\tt addrmf} and {\tt addarf} routines in {\tt FTOOLS}, weighed by the relative contribution of each observation to the total exposure time.  Spectra and responses from the two MOS detectors were co-added to create a single combined MOS spectrum, while spectra and responses extracted from the PN detector were summed separately.

The stacked MOS and PN spectra are fit with {\tt XSPEC} \citep[version 12.8.1p, ][]{xspec}.  The X-ray emission from the Galactic center region within the {\em XMM} field-of-view (radius of $\sim15'$) is a complicated combination of numerous sources, including active stars, cataclysmic variables, low and high-mass X-ray binaries, supernova remnants, thermal gas, particle and instrumental backgrounds.  As we are primarily interested in determining if excess line emission is present around 3.5 keV, we do not attempt to fit a physically meaningful spectral model that accounts for all of these components.  Instead, we focus on obtaining a good fit to the continuum emission in the spectral range of interest.  Specifically, we fit a narrow energy range from 2.3-4.5 keV.  This energy range is large enough to be much broader than the energy resolution of the detectors ($\sim 100$ eV) as well as containing several strong plasma emission lines based on which we can estimate the flux of weaker lines; at the same time, the energy window is small enough that we obtain a good fit with a simple power law continuum plus astrophysical line emission.  

\begin{figure*}
\begin{centering}
	\includegraphics[width=1.0\columnwidth]{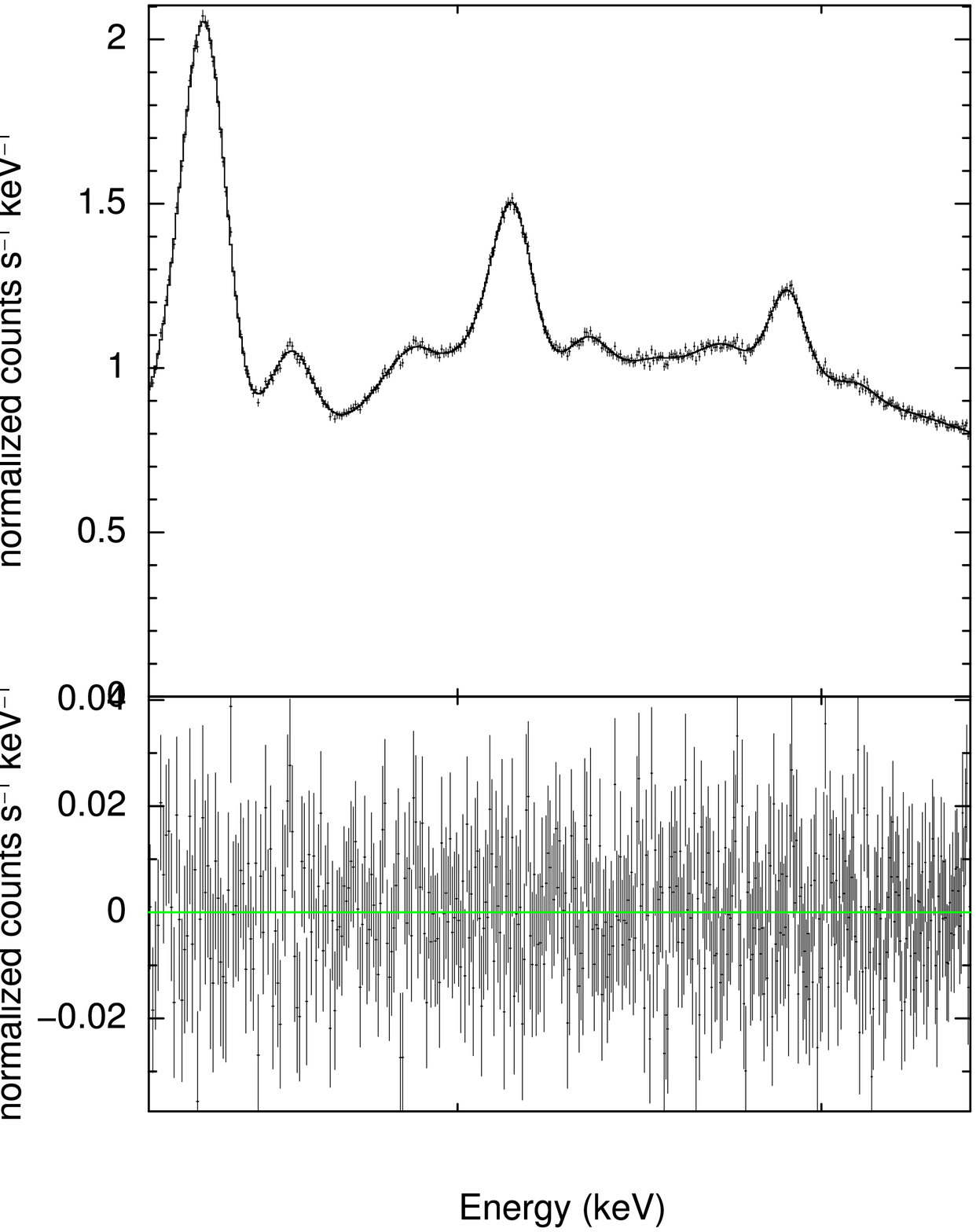}
	\includegraphics[width=1.0\columnwidth]{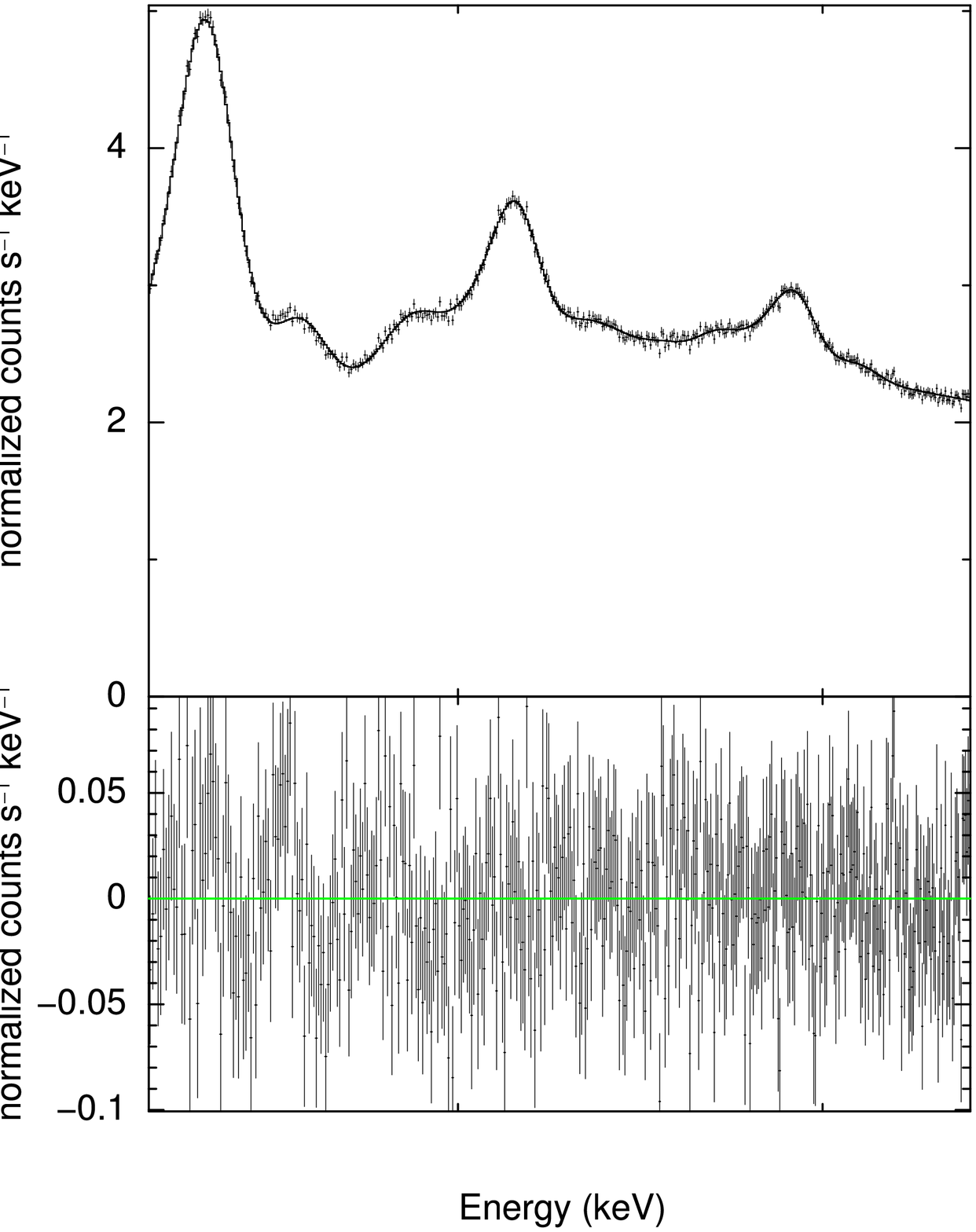}
	\end{centering}
\caption{ Stacked spectra of the Galactic center in the 2.3 to 4.5 keV band for the combined MOS (left) and PN (right) observations.  Also shown are the best-fit model and residuals.}
\label{fig:spectra}
\end{figure*}

We began by fitting the spectra to a simple power law plus a series of Gaussian lines modulated by photo-electric absorption.  Lines were added starting with the most significant plasma emission lines in the energy range taken from the {\tt AtomDB} database\footnote{http://www.atomdb.org/Webguide/webguide.php}\cite{apec}.  We then successively added Gaussians for weaker plasma lines.  Line energies were allowed to vary by $\pm 10$ eV to account for uncertainty in the energy calibration of the detectors and in the atomic database.  Known lines which improved the reduced $\chi^2$ were kept, while those that did not improve the fit and had very low normalizations were removed.  The lines included in the final fit (and the corresponding energies) were: Si XIII (2.34 keV), Si XIV (2.37, 2.50 keV), S XV (2.44, 2.46, 2.82, 2.88, 3.03 keV), S XVI (2.62, 3.39 keV), Ar XVII (3.13, 3.62, 3.69, 3.79, 4.0 keV), Ar XVIII (3.32 keV), K XVIII (3.48, 3.52 keV), Ca XIX (3.86, 3.90 keV), Ca XX (4.11 keV).  The resulting model yielded a very good fit to the data with a reduced $\chi^2$ of 0.96 ($\chi^2=388$/405 degrees of freedom) for the combined MOS spectrum and a reduced $\chi^2$ of 1.3 ($\chi^2=548$/410 degrees of freedom) for the combined PN spectrum.  Note that S XV emission around 2.45 keV results from a complex of several lines which are unresolved; here we model this emission as a set of two closely spaced Gaussians.  The remaining residuals in the PN fit are primarily due to modeling of this component. The best fit and residuals are shown in Figure~\ref{fig:spectra}, which illustrates that no additional excess line is present in the data beyond the lines listed above.

Note that there are two plasma lines due to K XVIII right around 3.5 keV.  Excluding these lines would significantly worsen the fit.  Clearly, it is of paramount importance to assess the expected strength of these plasma lines, as underestimating their strength could mimic a dark matter signal, while overestimating it could hide a true dark matter line.  In the following section, we take a similar approach to \citet{Bulbul:2014sua} by taking the measured strength of several strong plasma lines near 3.5 keV and using these, in conjunction with a variety of multi-temperature models, to estimate the strength of K XVIII emission.  Notice that there is also a pair of lines from Cl XVII at 3.51 keV which  are expected to contribute comparable flux to K XVIII for higher temperature plasmas ($kT \geq$ 5 keV).  These lines were not included in the analysis of \citet{Bulbul:2014sua}, but we will include here their estimated contribution, including constraints on the maximal brightness of the Cl XVII  Lyman-$\beta$ lines at 3.51 keV from the brightness of the corresponding Cl XVII Lyman-$\alpha$ at 2.96 keV.

\subsection{Predicting the Flux of the K XVIII Lines}\label{sec:plasma}

In Table~\ref{lineflux}, we list the measured fluxes of several strong emission lines in the spectral energy range under scrutiny as well as the sum of the two K XVIII lines.  When predicting emissivities, each of these lines is considered as the sum of all closely spaced, significant lines which are unresolvable at the instrumental energy resolution.  The lines/line complexes considered are: S XV (2.43 to 2.46 keV), S XVI (2.620+2.623 keV), Ar XVII (3.104+3.124+3.126+3.140 keV), K XVIII (3.476+3.515 keV), Ar XVIII (3.318+3.323 keV), Ca XIX (3.861+3.883+3.888+3.902 keV), and Ca XX (4.100+4.107 keV).  Note that we include the K XVIII lines at 3.476  keV and 3.515 keV separately in the spectral fit, but since the two normalizations are not independent, we sum them here. 
\begin{table*}
\centering
\begin{tabular}{cccc} \hline
Line &   Energy   &        MOS Flux  &  PN Flux\\
  & keV & photons cm$^{-2}$ s$^{-1}$ & photons cm$^{-2}$ s$^{-1}$\\
\hline
S XV &  2.45 & $3.1\pm0.2 \times 10^{-3}$ & $2.55\pm0.04 \times 10^{-3}$\\
S XVI	  &   2.62	&  $4.8\pm0.3 \times 10^{-4}$    & $2.9\pm0.1 \times 10^{-4}$ \\
Ar XVII   &    3.13    &   $6.1\pm0.1 \times 10^{-4}$   & $5.74\pm0.07 \times 10^{-4}$\\
Ar XVIII    &   3.32	&   $1.16\pm0.05 \times 10^{-4}$ & $6.5\pm1.0 \times 10^{-5}$\\
Ca XIX	&     3.90	 &  $2.55\pm0.03 \times 10^{-4}$  &  $2.4\pm0.1 \times 10^{-4}$\\
Ca XX &   4.1	 &  $4.1\pm0.3 \times 10^{-5}$  & $4.2\pm0.4 \times 10^{-5}$ \\
K XVIII (?) & 3.5 &  $4.5\pm0.4 \times 10^{-5}$ & $3.9\pm0.7 \times 10^{-5}$\\
\hline
\end{tabular}
\caption{Measured fluxes of the most prominent plasma lines between 2.4 and 4.5 keV in the Galactic center.  Fluxes are listed separately for the combined MOS and PN spectra.  Also listed is the summed flux of detected line emission at 3.48 and 3.52 keV, the position of plasma lines from K XVIII.}
\label{lineflux}
\end{table*}

In principle, the expected flux of K XVIII can be calculated based on the ratio of the emissivity of the K XVIII lines to these strong lines and the measured strong line fluxes.  However, the relative emissivities of different lines is a sensitive function of the plasma temperature.  In addition, the relative fluxes of lines of different elements depend on their relative abundances in the medium.  We will nominally assume that the relative elemental abundances track their ratios in the Sun \citep{1989GeCoA..53..197A}, though variation in the relative abundances by a factor of 2-3 would not be unreasonable \cite[e.g.][]{2004ApJ...613..326M,2005ApJ...631..964P, 2004MNRAS.350..129S, 2013PASJ...65...19U}.

Thermal emission from the Galactic center region is multi-temperature and is typically modeled as having contributions from both low temperature, $kT \sim 0.8-1$ keV, and high-temperature, $kT \sim 6-8$ keV, components \cite[e.g.][]{2004ApJ...613..326M, 2013MNRAS.434.1339H, 2013MNRAS.428.3462H, 2013PASJ...65...19U}. A portion of the high-temperature component may also stem from unresolved point sources \cite[e.g.][]{2002Natur.415..148W, 2009Natur.458.1142R, 2013MNRAS.428.3462H}.  Additional thermal components are contributed by individual sources in the region, including the Sagittarius A East supernova remnant \cite[e.g.][]{2005ApJ...631..964P, 2007PASJ...59S.237K} and the Arches massive star cluster \citep{2006MNRAS.371...38W, 2007PASJ...59S.229T}.  

Extrapolating a sensible multi-temperature model for this region is hindered by the unknown relative elemental abundances (which, as stated above, we putatively assume to track solar abundances). As a first guideline, we studied same-element ratios, which are not plagued by this systematic uncertainty, and which can be sensitive functions of temperature. We examined in particular three such ratios: (i) Ca XX to Ca XIX, (ii) Ar XVIII (3.32) to Ar XVII (3.13), and (iii) a set of S ratios. 
The measured Ca XX to Ca XIX ratio we observe indicates a temperature of about 2 keV; the Ar XVIII (3.32) to At XVII (3.13) ratio indicates a temperature of about 1.7 keV. Finally, we employed the ratio of a complex of S XV lines with energies in the vicinity of 2.4 keV, and of the S XVI  2.63 keV line. This ratio indicates a temperature of about 4.7 keV, but is potentially plagued by a steep dependence of the line fluxes on absorption, especially for the S XV low-energy lines.
 
 While keeping the same-element ratios into consideration, we first consider a set of two two-temperature models, motivated in part by previous results on modeling the GC multi-temperature plasma \citep{2004ApJ...613..326M, 2013MNRAS.434.1339H, 2013MNRAS.428.3462H, 2013PASJ...65...19U}. Specifically, we employ a two-temperature model, inspired by the results of \citet{2013PASJ...65...19U}, with $T_1=1$ keV and $T_2=7$ keV and a relative normalization $N_1/N_2=4$. For this model, we predict both the correct Ar 3.13 to Ar 3.32 ratio and the Ca XIX to Ca XX ratio to better than a factor 2; the model implies S to be a factor of about 3 under-abundant, for both MOS and PN observations, compared to the abundances of Ca and Ar inferred from the Ca XIX and Ar 3.32 lines (such abundances are within 10\% of each other), and the K to be a factor 2.4 (2.2) over-abundant for MOS (PN), if one wishes to attribute the 3.5 keV line to K XVIII entirely.
 
 We also consider an alternate two-temperature model, motivated by the results of \citet{2004ApJ...613..326M}, with $T_1=0.8$ keV and $T_2=8$ keV and a relative normalization $N_1/N_2=3$. In this case, normalizing to the Ca abundance as inferred from Ca XIX, we find reasonable but not perfect agreement for the Ar and Ca line ratios, with comparable Ar and Ca abundances; S needs to be a factor 2 (3) under-abundant for MOS (PN), and K a factor 3 (2.7) over-abundant for MOS (PN) observations.
 
 Finally, we consider a three-temperature model with $T_1=0.8$ keV, $T_2=2$ keV and $T_3=8$ keV and relative normalizations $N_1/N_2=0.17$ and $N_3/N_2=0.075$. We fine-tuned this model to provide a perfect match for the Ca XIX to Ca XX ratio and for the Ar 3.13 to Ar 3.32 ratios for MOS observations (good agreement is in place also for PN observations for these ratios, to within a few 10\%). The relative Argon to Calcium abundance is 0.8, thus very close to solar, for both MOS and PN. The relative Calcium to Sulfur abundance in this model has Sulfur under-abundant by a factor 2.7 (4.1 for PN), and the relative Calcium to Potassium abundance gives Potassium overabundant by a factor 2.7 (2.5 for PN).
 
 To investigate the potential contribution of the complex of Cl XVII Lyman-$\beta$ lines at 3.51 keV, we searched for lines corresponding to the brighter Cl XVII Lyman-$\alpha$ lines at 2.96 keV. We identified a line at 3.00 keV, with a line flux in the combined MOS spectrum of $6.33\times 10^{-5}\ {\rm ph}\ {\rm cm}^{-2}\ {\rm s}^{-1}$. The line is presumably contaminated by relatively bright S XV lines, especially one at 3.03 keV. We calculated, utilizing AtomDB v2.0.2, the ratio, as a function of temperature, of the Cl XVII Lyman-$\alpha$ lines to the S XV lines in the proximity of 3 keV, and found that the S XV complex dominates for low temperatures, below roughly 2.5 keV, while at high temperature the Cl XVII Lyman-$\alpha$ line dominates, to the level of being a factor 10 brighter for temperatures of around 8 keV and above.
 
 The Cl XVII Lyman-$\alpha$ (2.96) to Lyman-$\beta$ (3.51) ratio is very large at low temperature, and drops to between 6 and 7 for temperatures above approximately 2 keV. The maximal contribution to the 3.5 keV line from the Cl XVII Lyman-$\beta$ lines therefore can only stem from a relatively high-temperature plasma component. Specifically, the absolute largest contribution to the 3.5 keV line is a flux of about $1.0\times 10^{-5}\ {\rm ph}\ {\rm cm}^{-2}\ {\rm s}^{-1}$ at temperatures larger than 8 keV; at a temperature of 2 keV, the maximal contribution to the 3.5 keV line from Cl is about $0.3\times 10^{-5}\ {\rm ph}\ {\rm cm}^{-2}\ {\rm s}^{-1}$. Given that in our multi-temperature models a high-temperature component has at most a normalization suppressed by a factor 3 compared to the low-temperature component, we can safely conclude that the maximal Cl contribution to the 3.5 keV line is around $0.3\times 10^{-5}\ {\rm ph}\ {\rm cm}^{-2}\ {\rm s}^{-1}$, i.e. at most 7\% of the observed flux at 3.5 keV.

 In summary, we consistently find that multi-temperature models predict S to be under abundant by factors close to 3 and K to be over abundant by a similar factor, if one wishes to attribute the 3.5 keV line to K entirely (as explained above, a contamination of Cl is not excluded but subdominant).

\subsection{Implications for Dark Matter Decay}\label{sec:dm}

\begin{figure*}
\begin{centering}
	\includegraphics[width=0.9\columnwidth]{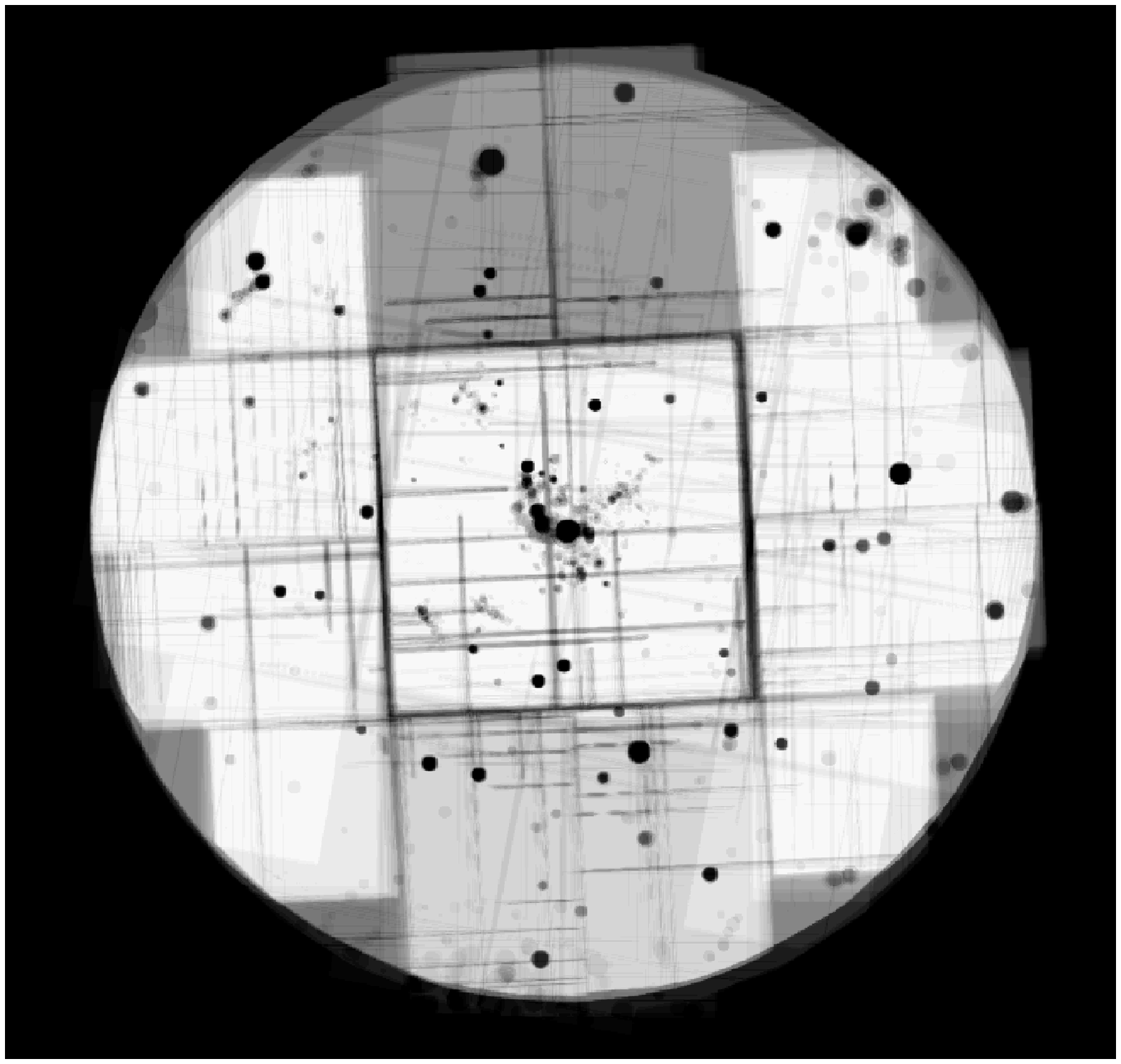}
	\includegraphics[width=0.9\columnwidth]{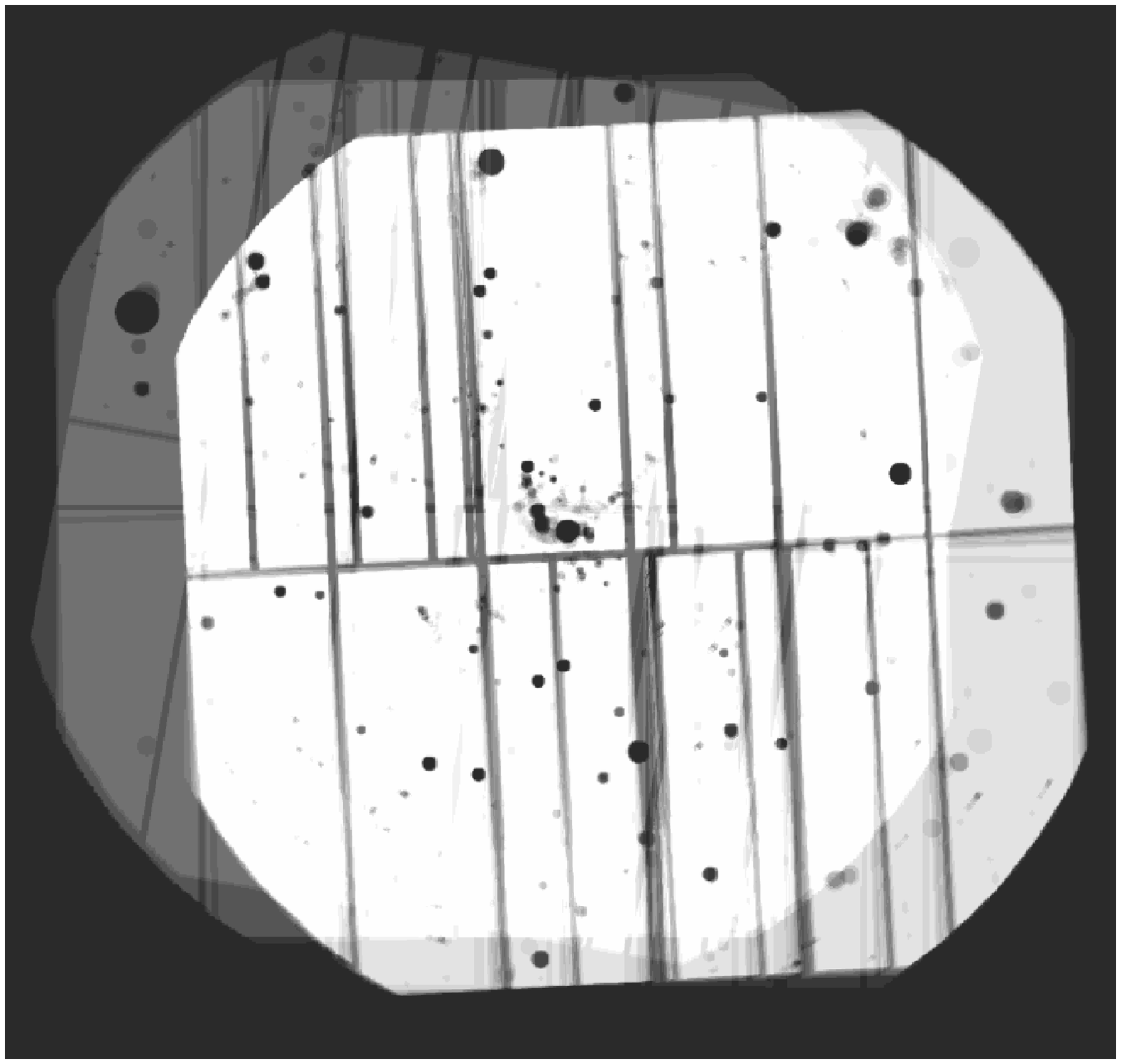}
	\end{centering}
\caption{Co-added images of the masks used in the MOS and PN spectral extraction produced by the {\tt ESAS} task {\tt cheese}.  Lost area due to point source exclusion and instrumental features are clear.  These summed masks are used to correct for the effective dark matter mass probed as a function of radius.}
\label{fig:mask}
\end{figure*}

In this section we describe the implications for a dark matter decay interpretation of the {\em XMM} observations analyzed above. We obtain two different results:

(i) a conservative constraint on the maximal dark matter lifetime for a dark matter particle decaying to a two-body final state including a photon of energy 3.5 keV, from a 2-sigma upper limit to the total X-ray flux at 3.5 keV from {\em XMM} MOS and PN observations, thus entirely neglecting any contribution from elemental lines; this is a conservative constraint in that even assuming solar abundances we find that at least of order  a third of the 3.5 keV emission comes from K XVIII;

(ii) a fit to the putative lifetime associated with a dark matter particle responsible for the observed 3.5 keV line flux, for a variety of dark matter density profiles.

We improve here on the calculation of similar constraints obtained by \citet{Riemer-Sorensen:2014yda} by considering the effective integrated line-of-sight integral of the dark matter density over the relevant angular region times the actual masks employed in the observations, and by taking into account the effect of absorption. Co-added images of the masks used in the MOS and PN spectral extraction are shown in Fig.~\ref{fig:mask} showing area lost due to point source exclusion and instrumental features as well as the positional offset between different observations.  For reference, we utilize two dark matter density profiles identical to those used in \citet{Riemer-Sorensen:2014yda} (to which we refer the Reader for further details), namely a Navarro-Frenk and White (NFW), and an Einasto profile (EIN) with $\alpha=0.17$ both with scale radius $r_s=21$ kpc and a local dark matter density of 0.4 GeV/cm$^3$. In addition, to obtain an even more conservative estimate of the dark matter ``column density'', we also employ a Burkert (BUR) profile, with a scale radius of 6 kpc and an identical local dark matter density (it is trivial to rescale our results for a different local dark matter density). We calculate the following effective $J$-factors in the direction of the Galactic center as follows
\begin{equation}
J\equiv \int_{\Delta\Omega} \int_{l.o.s.}A(\psi)\ \rho(\psi,l){\rm d}l\ {\rm d}\Omega(\psi)
\end{equation}
with $0\le A(\psi)\le 1$ the mask profile averaged over an annulus at an angle $\psi$ around the Galactic center, and with an angular aperture $\Delta\Omega$ roughly corresponding to the {\em XMM} field of view, i.e. a half-aperture angle of about 15 arcmin. We found the following effective $J$ values, in units of $10^{18}$ GeV cm$^{-2}$ (the numbers in parenthesis neglect the efficiency factor, i.e. assume $A(\psi)=1$):
\begin{eqnarray*}
&&\nonumber \hspace*{-1.6cm} {\rm PN}: J_{\rm BUR}=3.91\ (5.04);\quad J_{\rm NFW}=12.5\ (16.1);  \quad J_{\rm EIN}=14.9\ (19.2)\\
&&\nonumber \hspace*{-1.6cm} {\rm MOS}: J_{\rm BUR}=3.77\ (5.04);\quad J_{\rm NFW}=12.2\ (16.1);  \quad J_{\rm EIN}=14.4\ (19.2)
\end{eqnarray*}

(i) The {\em conservative} upper limit corresponds to the maximal flux at 2-sigma at an energy of 3.5 keV.  The resulting upper limits on the inverse dark matter lifetime (i.e. on the decay width $\Gamma=1/\tau$) are as follows, in units of $10^{-28}\ {\rm s}^{-1}$:
\begin{eqnarray}
&&\nonumber {\rm PN}: {\rm BUR}=12.0;\ {\rm NFW}=3.73;\ {\rm EIN}=3.13\\
&&\nonumber {\rm MOS}: {\rm BUR}=12.4;\ {\rm NFW}=3.82;\ {\rm EIN}=3.24
\end{eqnarray}

(ii) The preferred range for the sterile neutrino mixing angle assuming the entire 3.5 keV line flux originates from the two-body decay of a 7 keV sterile neutrino is, for PN, $\sin^2(2\theta)\sim 1.0-3.8)\times 10^{-10}$ (the lower number corresponding to the EIN profile, the higher number to the BUR profile case) and for MOS of $\sin^2(2\theta)\sim 1.2-4.5)\times 10^{-10}$. It is intriguing that the range of mixing angles we find here aligns well with the preferred values in \citet{Bulbul:2014sua, Boyarsky:2014jta} to explain the 3.5 keV line observed from clusters and from M31 in terms of dark matter decay. Of course, the line energy found for the GC is somewhat offset from that found for clusters.  

\section{Comparison to Clusters and M~31}

In this section, we reconsider the evidence for excess line emission near 3.5 keV in other systems, namely stacked observations of clusters and observations of M~31, in light of systematic uncertainties associated with the emission strength of the K XVIII lines.

 \subsection{Perseus and Other Clusters}\label{sec:clusters}

\citet{Bulbul:2014sua} present evidence of excess line emission around 3.55 to 3.57 keV from the stacked {\em XMM} spectra of 73 clusters of galaxies as well as individually from Perseus and a stack of Centaurus+Ophiuchus+Coma.  They also analyze Chandra observations of the Virgo cluster but do not find excess line emission from this cluster.   \citet{Bulbul:2014sua} predict the expected K XVIII line fluxes based on the measured fluxes of S XVI (2.62 keV), Ca XIX (3.9 keV), and Ca XX (4.1 keV) and multi-temperature plasma models.  Specifically, they fit a multi-temperature thermal plasma model (using a line-free apec model in XSPEC) with 2-4 thermal components.  The expected flux of the K XVIII lines is then calculated as a sum over the expectations for the different fit temperatures weighted by the relative normalizations of the individual thermal components from the spectral fit.  They then  allow the K XVIII line normalizations to go a factor of three above this prediction to account for potential relative elemental abundance variation.  They conclude that the excess emission detected is a factor of 10-20 larger than the KXVIII flux predicted in this way.

While this is a possible procedure, it may not capture the true thermal complexity of a cool-core system with AGN feedback like Perseus, nor of a stack of 73 clusters each with their own different intra-cluster medium structure.  In fact, it is important to note that the temperatures derived even from the multi-temperature fit do not agree for the same objects observed with different instruments \citep[e.g. Perseus for {\em XMM} MOS versus Perseus for {\em XMM} PN, see ][Table 2]{Bulbul:2014sua}, which leads to significantly differing predictions for K XVIII.  Here, we study whether, under reasonable assumptions, the K XVIII lines fluxes in clusters might be stronger than those used by \citet{Bulbul:2014sua} based on the quoted line strengths for S XVI, Ca XIX, and Ca XX in Tables 2 and 7 of their paper.

Our central tenet is that  \citet{Bulbul:2014sua} base their predictions on multi-temperature models biased towards large cluster temperatures, at which the K XVIII emissivity is systematically suppressed. At lower, yet reasonable cluster temperatures we show that the detected 3.5 keV lines can consistently be explained by K XVIII, to within a factor 3 or less in the K overabundance compared to S or Ca\footnote{We note that the limits on the Ly-$\alpha$ 2.96 keV line presented in \citet{Bulbul:2014ala} constrain the contribution of Cl XVII to the 3.5 to a negligible level of at most a few percent.}.

Using AtomDB 2.0.2, we calculated, for each cluster sample in \citet{Bulbul:2014ala}, the overabundance of K needed relative to S and Ca to explain the central value of the 3.5 keV line with the 3.48 plus 3.51 keV lines from K XVIII. In practice, following \citet{Bulbul:2014ala}, we calculate the maximal predicted K XVIII line flux $\Gamma_K$ at a given temperature $T_\varepsilon$ as 
\begin{equation}
\Gamma_K(r,T_\varepsilon)=\Gamma_r\frac{\varepsilon_K(T_\varepsilon)}{\varepsilon_r(T_\varepsilon)}
\end{equation}
with $r=$S, Ca XIX and Ca XX and $\Gamma_r$ the maximal line flux (central value plus 2$\sigma$) associated with the given measured strong line as quoted in \citet{Bulbul:2014ala}. The K overabundance is then the ratio of the central value of the observed 3.5 keV line flux to the predicted $\Gamma_K(r,T_\varepsilon)$.

Utilizing the Ca XIX line as a predictor, we find that for temperatures, $T\lesssim 3$ keV, K need not be overabundant compared to Ca by more than factors of 3-5, with the exception of the ``full sample - PN'' case where at $T=3$ keV the K overabundance is about 6 but drops at lower temperatures. Similarly for temperatures less than $3-3.5$ keV the Ca XX line implies that K need not be overabundant by more than about 2; for temperatures less than $\sim$2.5 keV, the Ca XX line actually indicates a K {\em under}abundance.  The S XVI line predicts somewhat lower K fluxes, similar to what is seen for the GC. The S XVI line predicts a bright enough K XVIII line in all cases with K to S abundances between 1 and 3 for very low temperatures, around 0.7 keV.  For larger temperatures, e.g. $T\sim 3$ keV, the K to S abundance ranges between factors of 5 and 20. 

In summary, we find that as long as the clusters' multi-temperature plasma includes a significant enough component at a relatively low temperatures ($T\lesssim 3$ keV) then the predicted K XVIII line flux is within a factor of at most 3 of the observed 3.5 keV line.

We believe that the reason why \citet{Bulbul:2014sua} arrived at differing conclusions from what we outline here is that they employed multi-temperature models systematically biased towards large temperatures. Let us make a very definite example. For the full-sample, PN stacked spectra, \citet{Bulbul:2014sua} used a multi-temperature model whose lowest temperature component had a temperature of 5.9 keV, the other three components being at 6.1, 7.3 and 10.9 keV. These high temperatures are clearly at odds with the Ca XX to Ca XIX line ratio that \citet{Bulbul:2014sua} quotes. This ratio is especially compelling, as it is (i) obviously unaffected by relative elemental abundance uncertainties, and (ii) a strong function of temperature. The measured Ca XX to Ca XIX line ratio differs from the prediction from the 4 temperatures employed by factors between approximately 3 and 10. As a result, the K XVIII is {\em under-estimated} by factors between 4.3 (for $T=5.9$ keV) up to more than 13 (for $T=10.9$ keV). Notice that this particular example is not unique, in that while lower-temperature components are present in other samples, the relative normalization almost always largely suppresses them, affecting the general conclusion presented in \citet{Bulbul:2014sua} that the contribution from K XVIII to the unidentified line is negligible. Additionally, since the fits carried out in \citet{Bulbul:2014sua} employed a maximal K XVIII flux which is very likely significantly under-estimated, serious doubts are cast on the conclusions derived in that study based on their spectral analysis.

We note in particular that the temperatures implied by the Ca XX to Ca XIX line ratio systematically indicate temperatures below 3 keV (specifically, between approximately 1.6 and 2.9 keV) for all XMM stacked cluster observations in \citet{Bulbul:2014sua}, clearly indicating that a significant plasma component with $T\lesssim 3$ keV should be present. As a result, the implied K overabundance should be significantly lower than what is derived and quoted in that paper.

\citet{Bulbul:2014sua} used their predicted K XVIII fluxes to set the allowed range for these lines in their spectral fits.  In addition to the concern outlined above as to whether their allowed maximal values for the K XVIII fluxes were truly conservative, when searching for an excess \citet{Bulbul:2014sua} do not enforce conservative fluxes for the K XVIII lines and the normalization of the ``excess" line is likely highly correlated to the K XVIII flux: in other words, the additional line might well have artificially absorbed photons from the K lines.  What \citet{Bulbul:2014sua} do quote is that if they allow the nearby Ar VXII line at 3.62 keV to have an arbitrarily large flux, the need for a line near 3.57 keV is removed. Something similar might be at work for the K XVIII lines as well.

\subsection{A Reanalysis of the {\em XMM} Data for M~31}\label{sec:m31}

\citet{Boyarsky:2014jta} report the detection of an unidentified line in both the Andromeda Galaxy and in a separate analysis of the Perseus Cluster.  \citet{Boyarsky:2014jta} do not report line fluxes for the astrophysical plasma lines included in their analysis.  To assess the significance of possible line emission in M~31, we therefore reanalyzed the available {\em XMM} data.  In particular, we analyzed the same M~31 data set employed by \cite{Boyarsky:2014jta}, namely the EPIC MOS data for the observations listed in their Table 3 which were determined to have low levels of contamination from particle background flares.  Here, we consider only the 29 {\em XMM} pointings within 2' of the center of M~31.  Data reduction and spectral stacking followed the same methodology employed for the Galactic center analysis described in Section~\ref{subsec:xmm}.

\begin{figure*}
\begin{centering}
	\includegraphics[width=1.0\columnwidth]{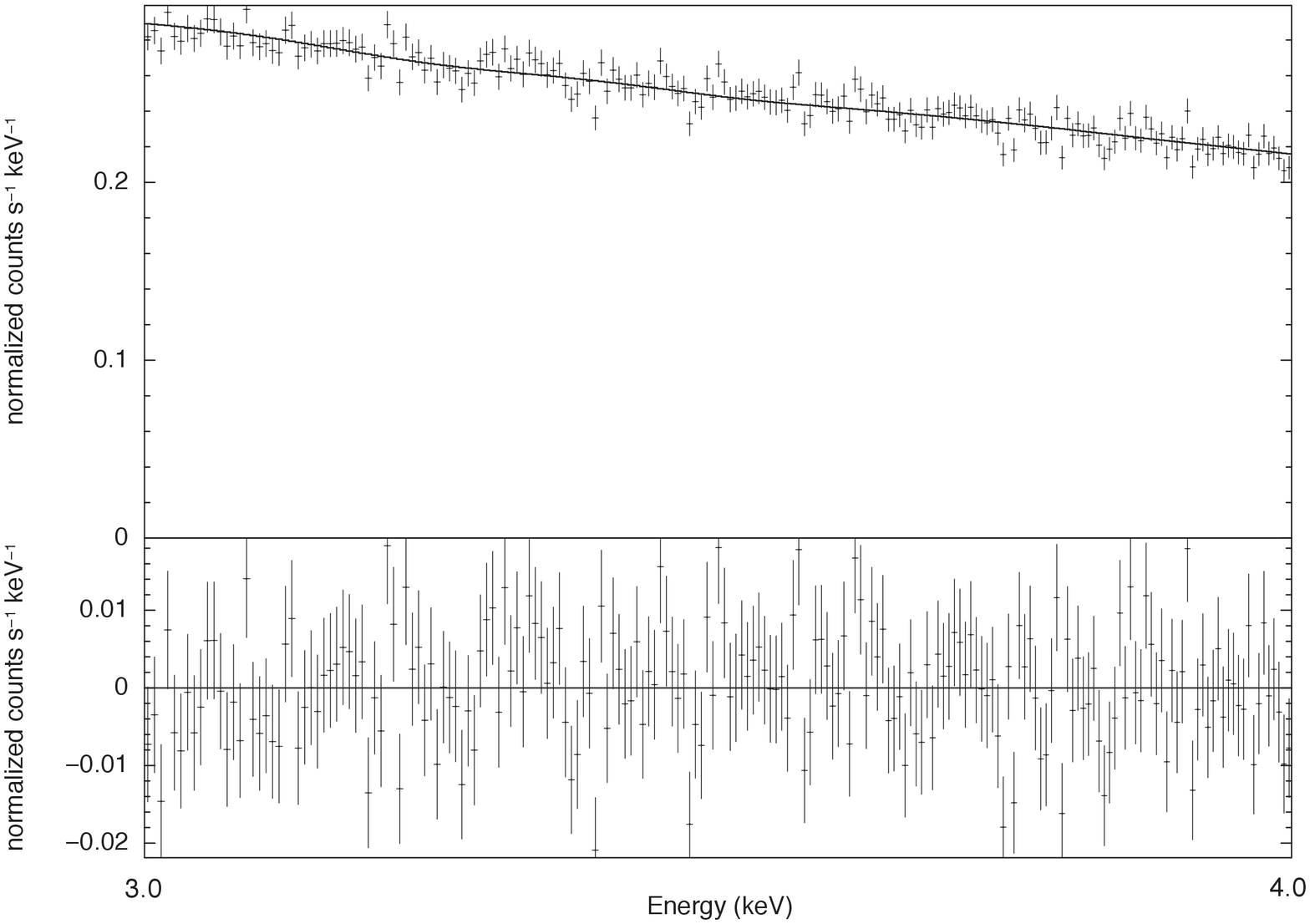}
	\includegraphics[width=1.0\columnwidth]{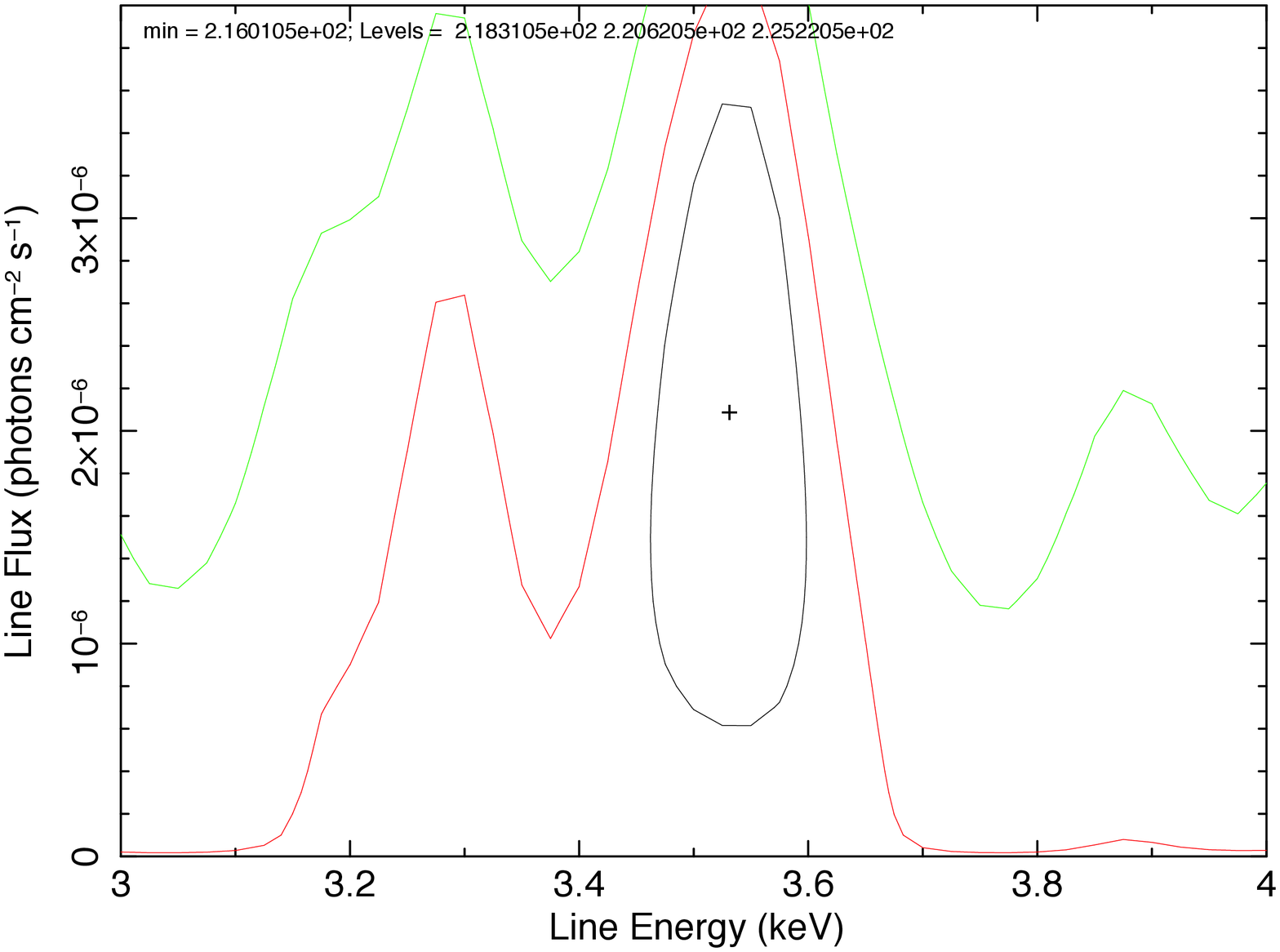}
	\end{centering}
\caption{ Left: Stacked MOS spectrum of M~31 in the 3.0 to 4.0 keV range along with the best-fit simple power law and residuals.  Right: Confidence contours on the combination of line energy and line flux after a Gaussian line is added to the spectrum on the left.  Contours show the 68\% (black), 90\% (red), and 99\% (green) confidence regions.}
\label{fig:m31}
\end{figure*}

Just as in the case of the Galactic center, the X-ray emission from M~31 is a complicated combination of sources, dominated by unresolved X-ray binaries and stellar sources in addition to thermal emission from a soft $\sim 0.3$ keV plasma \citep{2004ApJ...615..242T, 2008MNRAS.388...56B, 2010MNRAS.404.1879L}.  As with the Galactic center analysis, we concentrate on obtaining a good fit to the continuum emission near 3.5 keV rather than modeling in detail the contributions from each of these sources.  For the low average plasma temperature indicated, no strong line emission is expected in M~31 in the 3-4 keV range, nor is any evident.  As shown in the left panel of Fig.~\ref{fig:m31}, we find that the M~31 data are well fit by a simple power law between 3 and 4 keV, with a reduced $\chi^2$ of 1.12 ($\chi^2=220.6$/197 degrees of freedom) for the combined MOS spectrum.  

No significant residuals are evident in the fit, but we test for the existence of excess line emission by adding a Gaussian line with an energy allowed to vary between 3 to 4 keV.  We find that the addition of a line does not significantly improve the fit giving a reduced $\chi^2$ of 1.11 ($\chi^2=216.0$/195 degrees of freedom).  While the best-fit line energy is 3.53 keV as found by \cite{Boyarsky:2014jta}, the energy is essentially unconstrained and the line normalization is not significantly non-zero.  To illustrate this important point, the right panel of Fig.~\ref{fig:m31} shows confidence contours for the line energy and flux with contours indicating the 68\% (black), 90\% (red), and 99\% (green) confidence regions.   As can be seen from this figure, the line is only present at about the 1-sigma level and the normalization is consistent with zero with the 90\% confidence region.  The best-fit flux we derive is also a factor of two lower than what was found by \cite{Boyarsky:2014jta} whose flux may have been amplified by residuals in the spectrum outside of the 3-4 keV range.  

Extending the energy range to consider the data between 3-5 keV or 3-7 keV likewise does not lead to the detection of a line near 3.5 keV at more than 2-sigma significance \citep{tjspreply}.  For a broad energy range, the continuum cannot be model simply as a single power law; fitting energies above 5 keV our background model includes instrumental and astrophysical lines from Cr, Mn, and Fe as well as an additional unfolded power law to account for the particle background.  In the very broad, 2-8 keV energy range considered by \cite{Boyarsky:2014jta} additional astrophysical plasma lines and instrumental features also come into play.  Fitting this energy range with a model similar to \cite{Boyarsky:2014jta}, we find that spurious residuals appear in the spectrum near 3 keV which are not line-like in nature \cite[see e.g. Figure 2 in][]{reply}.  The most significant feature in these residuals is a {\em deficit} compared to the model between 2.8-3.1 keV followed by a slight excess extending from 3.1-3.8 keV.  The 3.5 keV ``excess" therefore results from poor continuum modeling.  We conclude that {\em no significant line emission near 3.5 keV is detected in M~31}.

\section{Discussion of Systematic Effects}\label{sec:sys}

There are three general possibilities for the origin of the 3.5 keV line seen in the GC and in clusters: (1) some form of new physics, such as dark matter decay or annihilation, or axion-photon conversion; (2) emission from potassium with either an abundance somewhat higher than nominally expected given our understanding of relative elemental abundances in the Sun, or with an emissivity higher than what predicted from AtomDB; or (3) systematics in the analysis and/or instrumental response.  In this section we review and address possible systematic effects.

Given the weakness of the 3.5 keV line, it is worth considering whether the line could originate from an instrumental feature such as a systematic error in the calibration of the instrumental response or from systematics in the analysis procedure.  For example, \cite{2014arXiv1412.1869T} find significant systematic errors in the the effective area calibration of Suzaku XIS using Crab Nebula observations, and argue that these might contribute to the detection of a line at 3.5 keV.  A 3.5 keV line has not been detected in XMM black fields, stacked dwarf, or stacked galaxy observations \citep{Boyarsky:2014jta, Malyshev:2014xqa, Anderson:2014tza}, arguing against an instrumental feature, which would presumably be present in all XMM observations.  In addition, as \cite{Bulbul:2014sua} stack clusters at a range of redshifts after shifting the spectra to the rest frame, instrumental features would be smeared out in their analysis, though it is not clear to what extent a few clusters dominate the 3.5 keV signal  in their stacks.  However, it is also possible that the analysis procedure itself leads to the detection of a spurious line or to overestimating the flux of a weak line.  \cite{2014arXiv1412.1869T} argued that the inclusion of several weak lines which significantly overlap given the instrumental energy resolution would lead to a suppressed continuum flux determination, creating artificial excesses in line-free regions (see their Figure 13).  Removing the plasma lines neighboring the 3.5 keV line in our GC fits significantly worsens the fit quality and does not remove the preference for a 3.5 keV line, but it remains possible that the presence of a large number of overlapping plasma lines could affect the flux determination of  weak lines.

As an additional test of possible systematic effects, we analyzed XMM observations of the Tycho supernova remnant (SNR), an object for which a new physics origin for a line at 3.5 keV similar to the ones advocated to explain the 3.5 keV line from clusters and galaxies would be highly unlikely.  Tycho shows overall similar plasma emission features \citep{2010ApJ...725..894H} to those seen in the GC.  Tycho was observed several times with XMM and we analyze the seven relatively flare-free observations (obsIDs 0096210101, 0310590101, 0310590201, 0412380201, 0412380301, 0412380401, 0511180101).  The data were reduced following that same procedure outlined in Section 2.1, concentrating on the combined MOS1 and MOS2 data.  The net flare free exposure time was 173 ksec for MOS1 and 176 ksec for MOS2.  Fitting the combined MOS spectrum extracted from the full FOV in the 2.3-4.5 keV range, strong emission lines due to S, Ar, and Ca are found, similar to those detected for the GC, with the exception of Ca XX at 4.1 keV which is not detected in Tycho.  These lines are velocity-broadened by 20-40 eV \citep{2010ApJ...725..894H}, and we fit for the line width using the brightest lines while constraining weaker lines of the same element to have the same width.  

As in the GC and clusters, we find that {\em also for the Tycho SNR the addition of a line near 3.5 keV significantly improves the fit}.  In this case, the best-fit line energy is 3.55 keV with a flux of $2.2\pm0.3 \times 10^{-5}$ photons cm$^{-2}$ s$^{-1}$.  The ratio of the flux of this line compared to the S XVI line flux is 0.08. If interpreted as emission from K XVIII despite the offset in energy, this flux ratio implies an overabundance of K relative to S compared to solar for any plasma temperature, similar to what is seen in the GC. This is completely at odds with theoretical predictions of elemental abundances from Type Ia supernovae, where normalized to solar abundances S is always significantly overabundant compared to K \citep[see e.g.][]{1984ApJ...286..644N,1999ApJS..125..439I, 2013MNRAS.429.1156S}.  

The comparison to other elements is strongly temperature dependent.  We note that the non-detection of Ca XX compared to the strong Ca XIX emission and the S XV to S XVI line ratio imply a low average plasma temperature less than $\sim 1$ keV.  For these temperatures, the ratio of the ArXVII flux to the 3.5 keV flux (ratio=0.026) likewise implies somewhat overabundant K, though for temperatures above 1 keV the ratio would be reasonable.  The Ca XIX flux matches reasonably well the 3.5 keV flux with a ratio of 0.034, with the K flux if anything somewhat low (by a factor of two) compared to solar ratios for low plasma temperatures.  We note that the above discussion assumes solar abundance ratios, while theoretical models of abundance yield from Type Ia supernovae imply that the relative K abundance compared to solar abundances should be suppressed with respect to Ca, Ar and S by factors of 3-10 or larger \citep{1984ApJ...286..644N, 1999ApJS..125..439I,  2004A&A...425.1029T,2010ApJ...712..624M, 2013MNRAS.429.1156S}.

The detection of 3.5 keV emission in Tycho argues for either systematic errors in the measurement and detection of weak lines or K lines brighter than nominally predicted, but certainly does not support a dark matter interpretation.

\section{Discussion and Conclusions}\label{sec:conclusions}

Reports of the detection of a 3.5 keV line from observations of clusters of galaxies and of M~31 \citep{Bulbul:2014sua, Boyarsky:2014jta} with no identified astrophysical line emission counterparts have triggered significant excitement, and much work on model-building of dark matter particle models that could explain the emission. No evidence for such an excess line was subsequently found in the analysis of {\em Chandra} data from the center of the Galaxy reported in \cite{Riemer-Sorensen:2014yda}, which also implied limits on a simple dark matter decay interpretation in tension with the preferred values inferred from clusters and from M31.

In this study, we analyzed {\em XMM} archival data from the Galactic center, with an effective total exposure about a factor 3 larger than the previous {\em Chandra} analysis.  We also carefully assessed the expected emissivity of astrophysical lines that could produce the observed 3.5 keV feature, associated with Potassium and Chlorine atomic transitions.  

We obtained an excellent fit to the {\em XMM} data 
by adding relevant plasma lines, including in particular two K XVIII lines at 3.48 and 3.52 keV explaining the observed 3.5 keV feature. Using the measured flux of brighter lines, we estimated a reasonable range for the flux of the K XVIII lines, and we found that the level of emission needed to fit the {\em XMM} data falls 
within the expected range. We thus found no indications of line emission near 3.5 keV in excess of what expected from known astrophysical plasma lines.

We then re-examined the possible role of Potassium and Chlorine lines in the cluster analysis of \citet{Bulbul:2014sua}, and found that inferring the emissivity of those lines from other measured lines, and for a reasonable range of temperatures, the 3.5 keV line can be  explained and no excess is clearly present. Finally, we re-analyzed {\em XMM} data from M~31, and showed that, in the relevant energy range, the spectrum is well fit by a featureless power law. We find no preference for a line at statistical significance greater than one sigma. The one-sigma excess we do find at energies around 3.5 keV also possesses a flux lower by about a factor of two compared to what claimed in \citet{Boyarsky:2014jta}.

We addressed possible systematic effects; while instrumental features seem unlikely, the possibility that the presence of a large number of tenuous overlapping plasma lines could affect the flux determination of the 3.5 keV line remains \citep{2014arXiv1412.1869T}. For example, we analyzed XMM observations of the Tycho SNR and found evidence for a 3.55 keV line which, if associated with potassium emission would imply, similar to the Galactic center results, a significant overabundance of potassium over, for example, sulfur. Since no new physics would contribute to a line from a SNR, and since the thermal emission observed from Tycho matches closely what seen from the Galactic center, this additional observation argues against a new physics interpretation of the 3.5 keV line.

In conclusion, while we do find evidence for a 3.5 keV line in X-ray data from the Galactic center, we showed that within the systematic uncertainty in the expected flux from known plasma lines, and considering additional uncertainty due to potential variation in the abundances of different elements, no conclusive excess line emission is present either from the Milky Way center or from clusters; also, no evidence was found of any statistically significant line from M~31  in the energy range of interest.

\section*{Acknowledgments}
%
\noindent We would like thank to Referee for their thoughtful comments on our manuscript.  We also thank A. ~Boyarsky and A.~Foster discussions of this work.  SP is partly supported by the US Department of Energy, Contract DE-SC0010107-001. 
%

\bibliography{galcenter}

\begin{thebibliography}{78}
\expandafter\ifx\csname natexlab\endcsname\relax\def\natexlab#1{#1}\fi
\expandafter\ifx\csname bibnamefont\endcsname\relax
  \def\bibnamefont#1{#1}\fi
\expandafter\ifx\csname bibfnamefont\endcsname\relax
  \def\bibfnamefont#1{#1}\fi
\expandafter\ifx\csname citenamefont\endcsname\relax
  \def\citenamefont#1{#1}\fi
\expandafter\ifx\csname url\endcsname\relax
  \def\url#1{\texttt{#1}}\fi
\expandafter\ifx\csname urlprefix\endcsname\relax\def\urlprefix{URL }\fi
\providecommand{\bibinfo}[2]{#2}
\providecommand{\eprint}[2][]{\url{#2}}

\bibitem[{\citenamefont{Boyarsky et~al.}(2009)\citenamefont{Boyarsky,
  Ruchayskiy, and Shaposhnikov}}]{Boyarsky:2009ix}
\bibinfo{author}{\bibfnamefont{A.}~\bibnamefont{Boyarsky}},
  \bibinfo{author}{\bibfnamefont{O.}~\bibnamefont{Ruchayskiy}},
  \bibnamefont{and}
  \bibinfo{author}{\bibfnamefont{M.}~\bibnamefont{Shaposhnikov}},
  \bibinfo{journal}{Ann.Rev.Nucl.Part.Sci.} \textbf{\bibinfo{volume}{59}},
  \bibinfo{pages}{191} (\bibinfo{year}{2009}), \eprint{0901.0011}.

\bibitem[{\citenamefont{Pal and Wolfenstein}(1982)}]{Pal:1981rm}
\bibinfo{author}{\bibfnamefont{P.~B.} \bibnamefont{Pal}} \bibnamefont{and}
  \bibinfo{author}{\bibfnamefont{L.}~\bibnamefont{Wolfenstein}},
  \bibinfo{journal}{Phys.Rev.} \textbf{\bibinfo{volume}{D25}},
  \bibinfo{pages}{766} (\bibinfo{year}{1982}).

\bibitem[{\citenamefont{Bulbul et~al.}(2014{\natexlab{a}})\citenamefont{Bulbul,
  Markevitch, Foster, Smith, Loewenstein et~al.}}]{Bulbul:2014sua}
\bibinfo{author}{\bibfnamefont{E.}~\bibnamefont{Bulbul}},
  \bibinfo{author}{\bibfnamefont{M.}~\bibnamefont{Markevitch}},
  \bibinfo{author}{\bibfnamefont{A.}~\bibnamefont{Foster}},
  \bibinfo{author}{\bibfnamefont{R.~K.} \bibnamefont{Smith}},
  \bibinfo{author}{\bibfnamefont{M.}~\bibnamefont{Loewenstein}},
  \bibnamefont{et~al.} (\bibinfo{year}{2014}{\natexlab{a}}),
  \eprint{1402.2301}.

\bibitem[{\citenamefont{Boyarsky et~al.}(2014)\citenamefont{Boyarsky,
  Ruchayskiy, Iakubovskyi, and Franse}}]{Boyarsky:2014jta}
\bibinfo{author}{\bibfnamefont{A.}~\bibnamefont{Boyarsky}},
  \bibinfo{author}{\bibfnamefont{O.}~\bibnamefont{Ruchayskiy}},
  \bibinfo{author}{\bibfnamefont{D.}~\bibnamefont{Iakubovskyi}},
  \bibnamefont{and} \bibinfo{author}{\bibfnamefont{J.}~\bibnamefont{Franse}}
  (\bibinfo{year}{2014}), \eprint{1402.4119}.

\bibitem[{\citenamefont{Ishida et~al.}(2014)\citenamefont{Ishida, Jeong, and
  Takahashi}}]{Ishida:2014dlp}
\bibinfo{author}{\bibfnamefont{H.}~\bibnamefont{Ishida}},
  \bibinfo{author}{\bibfnamefont{K.~S.} \bibnamefont{Jeong}}, \bibnamefont{and}
  \bibinfo{author}{\bibfnamefont{F.}~\bibnamefont{Takahashi}},
  \bibinfo{journal}{Phys.Lett.} \textbf{\bibinfo{volume}{B732}},
  \bibinfo{pages}{196} (\bibinfo{year}{2014}), \eprint{1402.5837}.

\bibitem[{\citenamefont{Abazajian}(2014)}]{Abazajian:2014gza}
\bibinfo{author}{\bibfnamefont{K.~N.} \bibnamefont{Abazajian}},
  \bibinfo{journal}{Phys.Rev.Lett.} \textbf{\bibinfo{volume}{112}},
  \bibinfo{pages}{161303} (\bibinfo{year}{2014}), \eprint{1403.0954}.

\bibitem[{\citenamefont{Baek and Okada}(2014)}]{Baek:2014qwa}
\bibinfo{author}{\bibfnamefont{S.}~\bibnamefont{Baek}} \bibnamefont{and}
  \bibinfo{author}{\bibfnamefont{H.}~\bibnamefont{Okada}}
  (\bibinfo{year}{2014}), \eprint{1403.1710}.

\bibitem[{\citenamefont{Tsuyuki}(2014)}]{Tsuyuki:2014aia}
\bibinfo{author}{\bibfnamefont{T.}~\bibnamefont{Tsuyuki}}
  (\bibinfo{year}{2014}), \eprint{1403.5053}.

\bibitem[{\citenamefont{Allahverdi et~al.}(2014)\citenamefont{Allahverdi,
  Dutta, and Gao}}]{Allahverdi:2014dqa}
\bibinfo{author}{\bibfnamefont{R.}~\bibnamefont{Allahverdi}},
  \bibinfo{author}{\bibfnamefont{B.}~\bibnamefont{Dutta}}, \bibnamefont{and}
  \bibinfo{author}{\bibfnamefont{Y.}~\bibnamefont{Gao}} (\bibinfo{year}{2014}),
  \eprint{1403.5717}.

\bibitem[{\citenamefont{Okada}(2014)}]{Okada:2014vla}
\bibinfo{author}{\bibfnamefont{H.}~\bibnamefont{Okada}} (\bibinfo{year}{2014}),
  \eprint{1404.0280}.

\bibitem[{\citenamefont{Modak}(2014)}]{Modak:2014vva}
\bibinfo{author}{\bibfnamefont{K.~P.} \bibnamefont{Modak}}
  (\bibinfo{year}{2014}), \eprint{1404.3676}.

\bibitem[{\citenamefont{Cline et~al.}(2014)\citenamefont{Cline, Farzan, Liu,
  Moore, and Xue}}]{Cline:2014eaa}
\bibinfo{author}{\bibfnamefont{J.~M.} \bibnamefont{Cline}},
  \bibinfo{author}{\bibfnamefont{Y.}~\bibnamefont{Farzan}},
  \bibinfo{author}{\bibfnamefont{Z.}~\bibnamefont{Liu}},
  \bibinfo{author}{\bibfnamefont{G.~D.} \bibnamefont{Moore}}, \bibnamefont{and}
  \bibinfo{author}{\bibfnamefont{W.}~\bibnamefont{Xue}} (\bibinfo{year}{2014}),
  \eprint{1404.3729}.

\bibitem[{\citenamefont{Rosner}(2014)}]{Rosner:2014ch}
\bibinfo{author}{\bibfnamefont{J.~L.} \bibnamefont{Rosner}}
  (\bibinfo{year}{2014}), \eprint{1404.5198}.

\bibitem[{\citenamefont{Robinson and Tsai}(2014)}]{Robinson:2014bma}
\bibinfo{author}{\bibfnamefont{D.~J.} \bibnamefont{Robinson}} \bibnamefont{and}
  \bibinfo{author}{\bibfnamefont{Y.}~\bibnamefont{Tsai}}
  (\bibinfo{year}{2014}), \eprint{1404.7118}.

\bibitem[{\citenamefont{Abada et~al.}(2014)\citenamefont{Abada, Arcadi, and
  Lucente}}]{Abada:2014zra}
\bibinfo{author}{\bibfnamefont{A.}~\bibnamefont{Abada}},
  \bibinfo{author}{\bibfnamefont{G.}~\bibnamefont{Arcadi}}, \bibnamefont{and}
  \bibinfo{author}{\bibfnamefont{M.}~\bibnamefont{Lucente}}
  (\bibinfo{year}{2014}), \eprint{1406.6556}.

\bibitem[{\citenamefont{Finkbeiner and Weiner}(2014)}]{Finkbeiner:2014sja}
\bibinfo{author}{\bibfnamefont{D.~P.} \bibnamefont{Finkbeiner}}
  \bibnamefont{and} \bibinfo{author}{\bibfnamefont{N.}~\bibnamefont{Weiner}}
  (\bibinfo{year}{2014}), \eprint{1402.6671}.

\bibitem[{\citenamefont{Okada and Toma}(2014)}]{Okada:2014zea}
\bibinfo{author}{\bibfnamefont{H.}~\bibnamefont{Okada}} \bibnamefont{and}
  \bibinfo{author}{\bibfnamefont{T.}~\bibnamefont{Toma}}
  (\bibinfo{year}{2014}), \eprint{1404.4795}.

\bibitem[{\citenamefont{Higaki et~al.}(2014)\citenamefont{Higaki, Jeong, and
  Takahashi}}]{Higaki:2014zua}
\bibinfo{author}{\bibfnamefont{T.}~\bibnamefont{Higaki}},
  \bibinfo{author}{\bibfnamefont{K.~S.} \bibnamefont{Jeong}}, \bibnamefont{and}
  \bibinfo{author}{\bibfnamefont{F.}~\bibnamefont{Takahashi}},
  \bibinfo{journal}{Phys.Lett.} \textbf{\bibinfo{volume}{B733}},
  \bibinfo{pages}{25} (\bibinfo{year}{2014}), \eprint{1402.6965}.

\bibitem[{\citenamefont{Jaeckel et~al.}(2014)\citenamefont{Jaeckel, Redondo,
  and Ringwald}}]{Jaeckel:2014qea}
\bibinfo{author}{\bibfnamefont{J.}~\bibnamefont{Jaeckel}},
  \bibinfo{author}{\bibfnamefont{J.}~\bibnamefont{Redondo}}, \bibnamefont{and}
  \bibinfo{author}{\bibfnamefont{A.}~\bibnamefont{Ringwald}},
  \bibinfo{journal}{Phys.Rev.} \textbf{\bibinfo{volume}{D89}},
  \bibinfo{pages}{103511} (\bibinfo{year}{2014}), \eprint{1402.7335}.

\bibitem[{\citenamefont{Lee et~al.}(2014)\citenamefont{Lee, Park, and
  Park}}]{Lee:2014xua}
\bibinfo{author}{\bibfnamefont{H.~M.} \bibnamefont{Lee}},
  \bibinfo{author}{\bibfnamefont{S.~C.} \bibnamefont{Park}}, \bibnamefont{and}
  \bibinfo{author}{\bibfnamefont{W.-I.} \bibnamefont{Park}}
  (\bibinfo{year}{2014}), \eprint{1403.0865}.

\bibitem[{\citenamefont{Cicoli et~al.}(2014)\citenamefont{Cicoli, Conlon,
  Marsh, and Rummel}}]{Cicoli:2014bfa}
\bibinfo{author}{\bibfnamefont{M.}~\bibnamefont{Cicoli}},
  \bibinfo{author}{\bibfnamefont{J.~P.} \bibnamefont{Conlon}},
  \bibinfo{author}{\bibfnamefont{M.~C.~D.} \bibnamefont{Marsh}},
  \bibnamefont{and} \bibinfo{author}{\bibfnamefont{M.}~\bibnamefont{Rummel}}
  (\bibinfo{year}{2014}), \eprint{1403.2370}.

\bibitem[{\citenamefont{Ringwald}(2014)}]{Ringwald:2014vqa}
\bibinfo{author}{\bibfnamefont{A.}~\bibnamefont{Ringwald}}
  (\bibinfo{year}{2014}), \eprint{1407.0546}.

\bibitem[{\citenamefont{Kong et~al.}(2014)\citenamefont{Kong, Park, and
  Park}}]{Kong:2014gea}
\bibinfo{author}{\bibfnamefont{K.}~\bibnamefont{Kong}},
  \bibinfo{author}{\bibfnamefont{J.-C.} \bibnamefont{Park}}, \bibnamefont{and}
  \bibinfo{author}{\bibfnamefont{S.~C.} \bibnamefont{Park}},
  \bibinfo{journal}{Phys.Lett.} \textbf{\bibinfo{volume}{B733}},
  \bibinfo{pages}{217} (\bibinfo{year}{2014}), \eprint{1403.1536}.

\bibitem[{\citenamefont{Choi and Seto}(2014)}]{Choi:2014tva}
\bibinfo{author}{\bibfnamefont{K.-Y.} \bibnamefont{Choi}} \bibnamefont{and}
  \bibinfo{author}{\bibfnamefont{O.}~\bibnamefont{Seto}}
  (\bibinfo{year}{2014}), \eprint{1403.1782}.

\bibitem[{\citenamefont{Dias et~al.}(2014)\citenamefont{Dias, Machado, Nishi,
  Ringwald, and Vaudrevange}}]{Dias:2014osa}
\bibinfo{author}{\bibfnamefont{A.}~\bibnamefont{Dias}},
  \bibinfo{author}{\bibfnamefont{A.}~\bibnamefont{Machado}},
  \bibinfo{author}{\bibfnamefont{C.}~\bibnamefont{Nishi}},
  \bibinfo{author}{\bibfnamefont{A.}~\bibnamefont{Ringwald}}, \bibnamefont{and}
  \bibinfo{author}{\bibfnamefont{P.}~\bibnamefont{Vaudrevange}}
  (\bibinfo{year}{2014}), \eprint{1403.5760}.

\bibitem[{\citenamefont{Liew}(2014)}]{Liew:2014gia}
\bibinfo{author}{\bibfnamefont{S.~P.} \bibnamefont{Liew}},
  \bibinfo{journal}{JCAP} \textbf{\bibinfo{volume}{05}}, \bibinfo{pages}{044}
  (\bibinfo{year}{2014}), \eprint{1403.6621}.

\bibitem[{\citenamefont{Conlon and Day}(2014)}]{Conlon:2014xsa}
\bibinfo{author}{\bibfnamefont{J.~P.} \bibnamefont{Conlon}} \bibnamefont{and}
  \bibinfo{author}{\bibfnamefont{F.~V.} \bibnamefont{Day}}
  (\bibinfo{year}{2014}), \eprint{1404.7741}.

\bibitem[{\citenamefont{Bomark and Roszkowski}(2014)}]{Bomark:2014yja}
\bibinfo{author}{\bibfnamefont{N.~E.} \bibnamefont{Bomark}} \bibnamefont{and}
  \bibinfo{author}{\bibfnamefont{L.}~\bibnamefont{Roszkowski}}
  (\bibinfo{year}{2014}), \eprint{1403.6503}.

\bibitem[{\citenamefont{Demidov and Gorbunov}(2014)}]{Demidov:2014hka}
\bibinfo{author}{\bibfnamefont{S.}~\bibnamefont{Demidov}} \bibnamefont{and}
  \bibinfo{author}{\bibfnamefont{D.}~\bibnamefont{Gorbunov}}
  (\bibinfo{year}{2014}), \eprint{1404.1339}.

\bibitem[{\citenamefont{Nakayama et~al.}(2014)\citenamefont{Nakayama,
  Takahashi, and Yanagida}}]{Nakayama:2014ova}
\bibinfo{author}{\bibfnamefont{K.}~\bibnamefont{Nakayama}},
  \bibinfo{author}{\bibfnamefont{F.}~\bibnamefont{Takahashi}},
  \bibnamefont{and} \bibinfo{author}{\bibfnamefont{T.~T.}
  \bibnamefont{Yanagida}} (\bibinfo{year}{2014}), \eprint{1403.1733}.

\bibitem[{\citenamefont{Shuve and Yavin}(2014)}]{Shuve:2014doa}
\bibinfo{author}{\bibfnamefont{B.}~\bibnamefont{Shuve}} \bibnamefont{and}
  \bibinfo{author}{\bibfnamefont{I.}~\bibnamefont{Yavin}}
  (\bibinfo{year}{2014}), \eprint{1403.2727}.

\bibitem[{\citenamefont{Kolda and Unwin}(2014)}]{Kolda:2014ppa}
\bibinfo{author}{\bibfnamefont{C.}~\bibnamefont{Kolda}} \bibnamefont{and}
  \bibinfo{author}{\bibfnamefont{J.}~\bibnamefont{Unwin}}
  (\bibinfo{year}{2014}), \eprint{1403.5580}.

\bibitem[{\citenamefont{Dutta et~al.}(2014)\citenamefont{Dutta, Gogoladze,
  Khalid, and Shafi}}]{Dutta:2014saa}
\bibinfo{author}{\bibfnamefont{B.}~\bibnamefont{Dutta}},
  \bibinfo{author}{\bibfnamefont{I.}~\bibnamefont{Gogoladze}},
  \bibinfo{author}{\bibfnamefont{R.}~\bibnamefont{Khalid}}, \bibnamefont{and}
  \bibinfo{author}{\bibfnamefont{Q.}~\bibnamefont{Shafi}}
  (\bibinfo{year}{2014}), \eprint{1407.0863}.

\bibitem[{\citenamefont{Baer et~al.}(2014)\citenamefont{Baer, Choi, Kim, and
  Roszkowski}}]{Baer:2014eja}
\bibinfo{author}{\bibfnamefont{H.}~\bibnamefont{Baer}},
  \bibinfo{author}{\bibfnamefont{K.-Y.} \bibnamefont{Choi}},
  \bibinfo{author}{\bibfnamefont{J.~E.} \bibnamefont{Kim}}, \bibnamefont{and}
  \bibinfo{author}{\bibfnamefont{L.}~\bibnamefont{Roszkowski}}
  (\bibinfo{year}{2014}), \eprint{1407.0017}.

\bibitem[{\citenamefont{Queiroz and Sinha}(2014)}]{Queiroz:2014yna}
\bibinfo{author}{\bibfnamefont{F.~S.} \bibnamefont{Queiroz}} \bibnamefont{and}
  \bibinfo{author}{\bibfnamefont{K.}~\bibnamefont{Sinha}}
  (\bibinfo{year}{2014}), \eprint{1404.1400}.

\bibitem[{\citenamefont{Lee}(2014)}]{Lee:2014koa}
\bibinfo{author}{\bibfnamefont{H.~M.} \bibnamefont{Lee}}
  (\bibinfo{year}{2014}), \eprint{1404.5446}.

\bibitem[{\citenamefont{Geng et~al.}(2014)\citenamefont{Geng, Huang, and
  Tsai}}]{Geng:2014zqa}
\bibinfo{author}{\bibfnamefont{C.-Q.} \bibnamefont{Geng}},
  \bibinfo{author}{\bibfnamefont{D.}~\bibnamefont{Huang}}, \bibnamefont{and}
  \bibinfo{author}{\bibfnamefont{L.-H.} \bibnamefont{Tsai}}
  (\bibinfo{year}{2014}), \eprint{1406.6481}.

\bibitem[{\citenamefont{Krall et~al.}(2014)\citenamefont{Krall, Reece, and
  Roxlo}}]{Krall:2014dba}
\bibinfo{author}{\bibfnamefont{R.}~\bibnamefont{Krall}},
  \bibinfo{author}{\bibfnamefont{M.}~\bibnamefont{Reece}}, \bibnamefont{and}
  \bibinfo{author}{\bibfnamefont{T.}~\bibnamefont{Roxlo}}
  (\bibinfo{year}{2014}), \eprint{1403.1240}.

\bibitem[{\citenamefont{Frandsen et~al.}(2014)\citenamefont{Frandsen, Sannino,
  Shoemaker, and Svendsen}}]{Frandsen:2014lfa}
\bibinfo{author}{\bibfnamefont{M.~T.} \bibnamefont{Frandsen}},
  \bibinfo{author}{\bibfnamefont{F.}~\bibnamefont{Sannino}},
  \bibinfo{author}{\bibfnamefont{I.~M.} \bibnamefont{Shoemaker}},
  \bibnamefont{and} \bibinfo{author}{\bibfnamefont{O.}~\bibnamefont{Svendsen}},
  \bibinfo{journal}{JCAP} \textbf{\bibinfo{volume}{1405}}, \bibinfo{pages}{033}
  (\bibinfo{year}{2014}), \eprint{1403.1570}.

\bibitem[{\citenamefont{Dudas et~al.}(2014)\citenamefont{Dudas, Heurtier, and
  Mambrini}}]{Dudas:2014ixa}
\bibinfo{author}{\bibfnamefont{E.}~\bibnamefont{Dudas}},
  \bibinfo{author}{\bibfnamefont{L.}~\bibnamefont{Heurtier}}, \bibnamefont{and}
  \bibinfo{author}{\bibfnamefont{Y.}~\bibnamefont{Mambrini}}
  (\bibinfo{year}{2014}), \eprint{1404.1927}.

\bibitem[{\citenamefont{Baek et~al.}(2014)\citenamefont{Baek, Ko, and
  Park}}]{Baek:2014poa}
\bibinfo{author}{\bibfnamefont{S.}~\bibnamefont{Baek}},
  \bibinfo{author}{\bibfnamefont{P.}~\bibnamefont{Ko}}, \bibnamefont{and}
  \bibinfo{author}{\bibfnamefont{W.-I.} \bibnamefont{Park}}
  (\bibinfo{year}{2014}), \eprint{1405.3730}.

\bibitem[{\citenamefont{Riemer-Sorensen}(2014)}]{Riemer-Sorensen:2014yda}
\bibinfo{author}{\bibfnamefont{S.}~\bibnamefont{Riemer-Sorensen}}
  (\bibinfo{year}{2014}), \eprint{1405.7943}.

\bibitem[{\citenamefont{Conlon and Powell}(2014)}]{Conlon:2014wna}
\bibinfo{author}{\bibfnamefont{J.~P.} \bibnamefont{Conlon}} \bibnamefont{and}
  \bibinfo{author}{\bibfnamefont{A.~J.} \bibnamefont{Powell}}
  (\bibinfo{year}{2014}), \eprint{1406.5518}.

\bibitem[{\citenamefont{Malyshev et~al.}(2014)\citenamefont{Malyshev, Neronov,
  and Eckert}}]{Malyshev:2014xqa}
\bibinfo{author}{\bibfnamefont{D.}~\bibnamefont{Malyshev}},
  \bibinfo{author}{\bibfnamefont{A.}~\bibnamefont{Neronov}}, \bibnamefont{and}
  \bibinfo{author}{\bibfnamefont{D.}~\bibnamefont{Eckert}}
  (\bibinfo{year}{2014}), \eprint{1408.3531}.

\bibitem[{\citenamefont{Anderson et~al.}(2014)\citenamefont{Anderson, Churazov,
  and Bregman}}]{Anderson:2014tza}
\bibinfo{author}{\bibfnamefont{M.~E.} \bibnamefont{Anderson}},
  \bibinfo{author}{\bibfnamefont{E.}~\bibnamefont{Churazov}}, \bibnamefont{and}
  \bibinfo{author}{\bibfnamefont{J.~N.} \bibnamefont{Bregman}}
  (\bibinfo{year}{2014}), \eprint{1408.4115}.

\bibitem[{\citenamefont{Urban et~al.}(2014)\citenamefont{Urban, Werner, Allen,
  Simionescu, Kaastra et~al.}}]{Urban:2014yda}
\bibinfo{author}{\bibfnamefont{O.}~\bibnamefont{Urban}},
  \bibinfo{author}{\bibfnamefont{N.}~\bibnamefont{Werner}},
  \bibinfo{author}{\bibfnamefont{S.}~\bibnamefont{Allen}},
  \bibinfo{author}{\bibfnamefont{A.}~\bibnamefont{Simionescu}},
  \bibinfo{author}{\bibfnamefont{J.}~\bibnamefont{Kaastra}},
  \bibnamefont{et~al.} (\bibinfo{year}{2014}), \eprint{1411.0050}.

\bibitem[{\citenamefont{Carlson et~al.}(2014)\citenamefont{Carlson, Jeltema,
  and Profumo}}]{Carlson:2014lla}
\bibinfo{author}{\bibfnamefont{E.}~\bibnamefont{Carlson}},
  \bibinfo{author}{\bibfnamefont{T.}~\bibnamefont{Jeltema}}, \bibnamefont{and}
  \bibinfo{author}{\bibfnamefont{S.}~\bibnamefont{Profumo}},
  \bibinfo{journal}{JCAP} \textbf{\bibinfo{volume}{xx}}, \bibinfo{pages}{nnn}
  (\bibinfo{year}{2014}), \eprint{1411.1758}.

\bibitem[{\citenamefont{{Snowden} et~al.}(2008)\citenamefont{{Snowden},
  {Mushotzky}, {Kuntz}, and {Davis}}}]{esas}
\bibinfo{author}{\bibfnamefont{S.~L.} \bibnamefont{{Snowden}}},
  \bibinfo{author}{\bibfnamefont{R.~F.} \bibnamefont{{Mushotzky}}},
  \bibinfo{author}{\bibfnamefont{K.~D.} \bibnamefont{{Kuntz}}},
  \bibnamefont{and} \bibinfo{author}{\bibfnamefont{D.~S.}
  \bibnamefont{{Davis}}}, \bibinfo{journal}{Astronomy \& Astrophysics}
  \textbf{\bibinfo{volume}{478}}, \bibinfo{pages}{615} (\bibinfo{year}{2008}),
  \eprint{0710.2241}.

\bibitem[{\citenamefont{{Kuntz} and {Snowden}}(2008)}]{esas2}
\bibinfo{author}{\bibfnamefont{K.~D.} \bibnamefont{{Kuntz}}} \bibnamefont{and}
  \bibinfo{author}{\bibfnamefont{S.~L.} \bibnamefont{{Snowden}}},
  \bibinfo{journal}{Astronomy \& Astrophysics} \textbf{\bibinfo{volume}{478}},
  \bibinfo{pages}{575} (\bibinfo{year}{2008}).

\bibitem[{\citenamefont{{Nowak} et~al.}(2012)\citenamefont{{Nowak}, {Neilsen},
  {Markoff}, {Baganoff}, {Porquet}, {Grosso}, {Levin}, {Houck}, {Eckart},
  {Falcke} et~al.}}]{2012ApJ...759...95N}
\bibinfo{author}{\bibfnamefont{M.~A.} \bibnamefont{{Nowak}}},
  \bibinfo{author}{\bibfnamefont{J.}~\bibnamefont{{Neilsen}}},
  \bibinfo{author}{\bibfnamefont{S.~B.} \bibnamefont{{Markoff}}},
  \bibinfo{author}{\bibfnamefont{F.~K.} \bibnamefont{{Baganoff}}},
  \bibinfo{author}{\bibfnamefont{D.}~\bibnamefont{{Porquet}}},
  \bibinfo{author}{\bibfnamefont{N.}~\bibnamefont{{Grosso}}},
  \bibinfo{author}{\bibfnamefont{Y.}~\bibnamefont{{Levin}}},
  \bibinfo{author}{\bibfnamefont{J.}~\bibnamefont{{Houck}}},
  \bibinfo{author}{\bibfnamefont{A.}~\bibnamefont{{Eckart}}},
  \bibinfo{author}{\bibfnamefont{H.}~\bibnamefont{{Falcke}}},
  \bibnamefont{et~al.}, \bibinfo{journal}{Astrophys.J.}
  \textbf{\bibinfo{volume}{759}}, \bibinfo{eid}{95} (\bibinfo{year}{2012}),
  \eprint{1209.6354}.

\bibitem[{\citenamefont{{Blackburn}}(1995)}]{ftools}
\bibinfo{author}{\bibfnamefont{J.~K.} \bibnamefont{{Blackburn}}}, in
  \emph{\bibinfo{booktitle}{Astronomical Data Analysis Software and Systems
  IV}}, edited by \bibinfo{editor}{\bibfnamefont{R.~A.} \bibnamefont{{Shaw}}},
  \bibinfo{editor}{\bibfnamefont{H.~E.} \bibnamefont{{Payne}}},
  \bibnamefont{and} \bibinfo{editor}{\bibfnamefont{J.~J.~E.}
  \bibnamefont{{Hayes}}} (\bibinfo{year}{1995}), vol.~\bibinfo{volume}{77} of
  \emph{\bibinfo{series}{Astronomical Society of the Pacific Conference
  Series}}, p. \bibinfo{pages}{367}.

\bibitem[{\citenamefont{{Arnaud}}(1996)}]{xspec}
\bibinfo{author}{\bibfnamefont{K.~A.} \bibnamefont{{Arnaud}}}, in
  \emph{\bibinfo{booktitle}{Astronomical Data Analysis Software and Systems
  V}}, edited by \bibinfo{editor}{\bibfnamefont{G.~H.} \bibnamefont{{Jacoby}}}
  \bibnamefont{and} \bibinfo{editor}{\bibfnamefont{J.}~\bibnamefont{{Barnes}}}
  (\bibinfo{year}{1996}), vol. \bibinfo{volume}{101} of
  \emph{\bibinfo{series}{Astronomical Society of the Pacific Conference
  Series}}, p.~\bibinfo{pages}{17}.

\bibitem[{\citenamefont{{Smith} et~al.}(2001)\citenamefont{{Smith},
  {Brickhouse}, {Liedahl}, and {Raymond}}}]{apec}
\bibinfo{author}{\bibfnamefont{R.~K.} \bibnamefont{{Smith}}},
  \bibinfo{author}{\bibfnamefont{N.~S.} \bibnamefont{{Brickhouse}}},
  \bibinfo{author}{\bibfnamefont{D.~A.} \bibnamefont{{Liedahl}}},
  \bibnamefont{and} \bibinfo{author}{\bibfnamefont{J.~C.}
  \bibnamefont{{Raymond}}}, \bibinfo{journal}{ApJ Letters}
  \textbf{\bibinfo{volume}{556}}, \bibinfo{pages}{L91} (\bibinfo{year}{2001}),
  \eprint{astro-ph/0106478}.

\bibitem[{\citenamefont{{Anders} and {Grevesse}}(1989)}]{1989GeCoA..53..197A}
\bibinfo{author}{\bibfnamefont{E.}~\bibnamefont{{Anders}}} \bibnamefont{and}
  \bibinfo{author}{\bibfnamefont{N.}~\bibnamefont{{Grevesse}}},
  \bibinfo{journal}{Geochimica et Cosmochimica Acta}
  \textbf{\bibinfo{volume}{53}}, \bibinfo{pages}{197} (\bibinfo{year}{1989}).

\bibitem[{\citenamefont{{Muno} et~al.}(2004)\citenamefont{{Muno}, {Baganoff},
  {Bautz}, {Feigelson}, {Garmire}, {Morris}, {Park}, {Ricker}, and
  {Townsley}}}]{2004ApJ...613..326M}
\bibinfo{author}{\bibfnamefont{M.~P.} \bibnamefont{{Muno}}},
  \bibinfo{author}{\bibfnamefont{F.~K.} \bibnamefont{{Baganoff}}},
  \bibinfo{author}{\bibfnamefont{M.~W.} \bibnamefont{{Bautz}}},
  \bibinfo{author}{\bibfnamefont{E.~D.} \bibnamefont{{Feigelson}}},
  \bibinfo{author}{\bibfnamefont{G.~P.} \bibnamefont{{Garmire}}},
  \bibinfo{author}{\bibfnamefont{M.~R.} \bibnamefont{{Morris}}},
  \bibinfo{author}{\bibfnamefont{S.}~\bibnamefont{{Park}}},
  \bibinfo{author}{\bibfnamefont{G.~R.} \bibnamefont{{Ricker}}},
  \bibnamefont{and} \bibinfo{author}{\bibfnamefont{L.~K.}
  \bibnamefont{{Townsley}}}, \bibinfo{journal}{Astrophys.J.}
  \textbf{\bibinfo{volume}{613}}, \bibinfo{pages}{326} (\bibinfo{year}{2004}),
  \eprint{astro-ph/0402087}.

\bibitem[{\citenamefont{{Park} et~al.}(2005)\citenamefont{{Park}, {Muno},
  {Baganoff}, {Maeda}, {Morris}, {Chartas}, {Sanwal}, {Burrows}, and
  {Garmire}}}]{2005ApJ...631..964P}
\bibinfo{author}{\bibfnamefont{S.}~\bibnamefont{{Park}}},
  \bibinfo{author}{\bibfnamefont{M.~P.} \bibnamefont{{Muno}}},
  \bibinfo{author}{\bibfnamefont{F.~K.} \bibnamefont{{Baganoff}}},
  \bibinfo{author}{\bibfnamefont{Y.}~\bibnamefont{{Maeda}}},
  \bibinfo{author}{\bibfnamefont{M.}~\bibnamefont{{Morris}}},
  \bibinfo{author}{\bibfnamefont{G.}~\bibnamefont{{Chartas}}},
  \bibinfo{author}{\bibfnamefont{D.}~\bibnamefont{{Sanwal}}},
  \bibinfo{author}{\bibfnamefont{D.~N.} \bibnamefont{{Burrows}}},
  \bibnamefont{and} \bibinfo{author}{\bibfnamefont{G.~P.}
  \bibnamefont{{Garmire}}}, \bibinfo{journal}{Astrophys.J.}
  \textbf{\bibinfo{volume}{631}}, \bibinfo{pages}{964} (\bibinfo{year}{2005}),
  \eprint{astro-ph/0506168}.

\bibitem[{\citenamefont{{Sakano} et~al.}(2004)\citenamefont{{Sakano},
  {Warwick}, {Decourchelle}, and {Predehl}}}]{2004MNRAS.350..129S}
\bibinfo{author}{\bibfnamefont{M.}~\bibnamefont{{Sakano}}},
  \bibinfo{author}{\bibfnamefont{R.~S.} \bibnamefont{{Warwick}}},
  \bibinfo{author}{\bibfnamefont{A.}~\bibnamefont{{Decourchelle}}},
  \bibnamefont{and}
  \bibinfo{author}{\bibfnamefont{P.}~\bibnamefont{{Predehl}}},
  \bibinfo{journal}{MNRAS} \textbf{\bibinfo{volume}{350}}, \bibinfo{pages}{129}
  (\bibinfo{year}{2004}), \eprint{astro-ph/0312541}.

\bibitem[{\citenamefont{{Uchiyama} et~al.}(2013)\citenamefont{{Uchiyama},
  {Nobukawa}, {Tsuru}, and {Koyama}}}]{2013PASJ...65...19U}
\bibinfo{author}{\bibfnamefont{H.}~\bibnamefont{{Uchiyama}}},
  \bibinfo{author}{\bibfnamefont{M.}~\bibnamefont{{Nobukawa}}},
  \bibinfo{author}{\bibfnamefont{T.~G.} \bibnamefont{{Tsuru}}},
  \bibnamefont{and} \bibinfo{author}{\bibfnamefont{K.}~\bibnamefont{{Koyama}}},
  \bibinfo{journal}{Publications of the Astronomical Society of Japan}
  \textbf{\bibinfo{volume}{65}}, \bibinfo{pages}{19} (\bibinfo{year}{2013}),
  \eprint{1209.0067}.

\bibitem[{\citenamefont{{Heard} and
  {Warwick}}(2013{\natexlab{a}})}]{2013MNRAS.434.1339H}
\bibinfo{author}{\bibfnamefont{V.}~\bibnamefont{{Heard}}} \bibnamefont{and}
  \bibinfo{author}{\bibfnamefont{R.~S.} \bibnamefont{{Warwick}}},
  \bibinfo{journal}{MNRAS} \textbf{\bibinfo{volume}{434}},
  \bibinfo{pages}{1339} (\bibinfo{year}{2013}{\natexlab{a}}),
  \eprint{1306.4186}.

\bibitem[{\citenamefont{{Heard} and
  {Warwick}}(2013{\natexlab{b}})}]{2013MNRAS.428.3462H}
\bibinfo{author}{\bibfnamefont{V.}~\bibnamefont{{Heard}}} \bibnamefont{and}
  \bibinfo{author}{\bibfnamefont{R.~S.} \bibnamefont{{Warwick}}},
  \bibinfo{journal}{MNRAS} \textbf{\bibinfo{volume}{428}},
  \bibinfo{pages}{3462} (\bibinfo{year}{2013}{\natexlab{b}}),
  \eprint{1210.6808}.

\bibitem[{\citenamefont{{Wang} et~al.}(2002)\citenamefont{{Wang}, {Gotthelf},
  and {Lang}}}]{2002Natur.415..148W}
\bibinfo{author}{\bibfnamefont{Q.~D.} \bibnamefont{{Wang}}},
  \bibinfo{author}{\bibfnamefont{E.~V.} \bibnamefont{{Gotthelf}}},
  \bibnamefont{and} \bibinfo{author}{\bibfnamefont{C.~C.}
  \bibnamefont{{Lang}}}, \bibinfo{journal}{Nature}
  \textbf{\bibinfo{volume}{415}}, \bibinfo{pages}{148} (\bibinfo{year}{2002}).

\bibitem[{\citenamefont{{Revnivtsev} et~al.}(2009)\citenamefont{{Revnivtsev},
  {Sazonov}, {Churazov}, {Forman}, {Vikhlinin}, and
  {Sunyaev}}}]{2009Natur.458.1142R}
\bibinfo{author}{\bibfnamefont{M.}~\bibnamefont{{Revnivtsev}}},
  \bibinfo{author}{\bibfnamefont{S.}~\bibnamefont{{Sazonov}}},
  \bibinfo{author}{\bibfnamefont{E.}~\bibnamefont{{Churazov}}},
  \bibinfo{author}{\bibfnamefont{W.}~\bibnamefont{{Forman}}},
  \bibinfo{author}{\bibfnamefont{A.}~\bibnamefont{{Vikhlinin}}},
  \bibnamefont{and}
  \bibinfo{author}{\bibfnamefont{R.}~\bibnamefont{{Sunyaev}}},
  \bibinfo{journal}{Nature} \textbf{\bibinfo{volume}{458}},
  \bibinfo{pages}{1142} (\bibinfo{year}{2009}), \eprint{0904.4649}.

\bibitem[{\citenamefont{{Koyama} et~al.}(2007)\citenamefont{{Koyama},
  {Uchiyama}, {Hyodo}, {Matsumoto}, {Tsuru}, {Ozaki}, {Maeda}, and
  {Murakami}}}]{2007PASJ...59S.237K}
\bibinfo{author}{\bibfnamefont{K.}~\bibnamefont{{Koyama}}},
  \bibinfo{author}{\bibfnamefont{H.}~\bibnamefont{{Uchiyama}}},
  \bibinfo{author}{\bibfnamefont{Y.}~\bibnamefont{{Hyodo}}},
  \bibinfo{author}{\bibfnamefont{H.}~\bibnamefont{{Matsumoto}}},
  \bibinfo{author}{\bibfnamefont{T.~G.} \bibnamefont{{Tsuru}}},
  \bibinfo{author}{\bibfnamefont{M.}~\bibnamefont{{Ozaki}}},
  \bibinfo{author}{\bibfnamefont{Y.}~\bibnamefont{{Maeda}}}, \bibnamefont{and}
  \bibinfo{author}{\bibfnamefont{H.}~\bibnamefont{{Murakami}}},
  \bibinfo{journal}{Publications of the Astronomical Society of Japan}
  \textbf{\bibinfo{volume}{59}}, \bibinfo{pages}{237} (\bibinfo{year}{2007}).

\bibitem[{\citenamefont{{Wang} et~al.}(2006)\citenamefont{{Wang}, {Dong}, and
  {Lang}}}]{2006MNRAS.371...38W}
\bibinfo{author}{\bibfnamefont{Q.~D.} \bibnamefont{{Wang}}},
  \bibinfo{author}{\bibfnamefont{H.}~\bibnamefont{{Dong}}}, \bibnamefont{and}
  \bibinfo{author}{\bibfnamefont{C.}~\bibnamefont{{Lang}}},
  \bibinfo{journal}{MNRAS} \textbf{\bibinfo{volume}{371}}, \bibinfo{pages}{38}
  (\bibinfo{year}{2006}), \eprint{astro-ph/0606282}.

\bibitem[{\citenamefont{{Tsujimoto} et~al.}(2007)\citenamefont{{Tsujimoto},
  {Hyodo}, and {Koyama}}}]{2007PASJ...59S.229T}
\bibinfo{author}{\bibfnamefont{M.}~\bibnamefont{{Tsujimoto}}},
  \bibinfo{author}{\bibfnamefont{Y.}~\bibnamefont{{Hyodo}}}, \bibnamefont{and}
  \bibinfo{author}{\bibfnamefont{K.}~\bibnamefont{{Koyama}}},
  \bibinfo{journal}{Publications of the Astronomical Society of Japan}
  \textbf{\bibinfo{volume}{59}}, \bibinfo{pages}{229} (\bibinfo{year}{2007}),
  \eprint{astro-ph/0611104}.

\bibitem[{\citenamefont{Bulbul et~al.}(2014{\natexlab{b}})\citenamefont{Bulbul,
  Markevitch, Foster, Smith, Loewenstein et~al.}}]{Bulbul:2014ala}
\bibinfo{author}{\bibfnamefont{E.}~\bibnamefont{Bulbul}},
  \bibinfo{author}{\bibfnamefont{M.}~\bibnamefont{Markevitch}},
  \bibinfo{author}{\bibfnamefont{A.~R.} \bibnamefont{Foster}},
  \bibinfo{author}{\bibfnamefont{R.~K.} \bibnamefont{Smith}},
  \bibinfo{author}{\bibfnamefont{M.}~\bibnamefont{Loewenstein}},
  \bibnamefont{et~al.} (\bibinfo{year}{2014}{\natexlab{b}}),
  \eprint{1409.4143}.

\bibitem[{\citenamefont{{Takahashi} et~al.}(2004)\citenamefont{{Takahashi},
  {Okada}, {Kokubun}, and {Makishima}}}]{2004ApJ...615..242T}
\bibinfo{author}{\bibfnamefont{H.}~\bibnamefont{{Takahashi}}},
  \bibinfo{author}{\bibfnamefont{Y.}~\bibnamefont{{Okada}}},
  \bibinfo{author}{\bibfnamefont{M.}~\bibnamefont{{Kokubun}}},
  \bibnamefont{and}
  \bibinfo{author}{\bibfnamefont{K.}~\bibnamefont{{Makishima}}},
  \bibinfo{journal}{Astrophys.J.} \textbf{\bibinfo{volume}{615}},
  \bibinfo{pages}{242} (\bibinfo{year}{2004}), \eprint{astro-ph/0408305}.

\bibitem[{\citenamefont{{Bogd{\'a}n} and
  {Gilfanov}}(2008)}]{2008MNRAS.388...56B}
\bibinfo{author}{\bibfnamefont{{\'A}.}~\bibnamefont{{Bogd{\'a}n}}}
  \bibnamefont{and}
  \bibinfo{author}{\bibfnamefont{M.}~\bibnamefont{{Gilfanov}}},
  \bibinfo{journal}{MNRAS} \textbf{\bibinfo{volume}{388}}, \bibinfo{pages}{56}
  (\bibinfo{year}{2008}), \eprint{0803.0063}.

\bibitem[{\citenamefont{{Liu} et~al.}(2010)\citenamefont{{Liu}, {Wang}, {Li},
  and {Peterson}}}]{2010MNRAS.404.1879L}
\bibinfo{author}{\bibfnamefont{J.}~\bibnamefont{{Liu}}},
  \bibinfo{author}{\bibfnamefont{Q.~D.} \bibnamefont{{Wang}}},
  \bibinfo{author}{\bibfnamefont{Z.}~\bibnamefont{{Li}}}, \bibnamefont{and}
  \bibinfo{author}{\bibfnamefont{J.~R.} \bibnamefont{{Peterson}}},
  \bibinfo{journal}{MNRAS} \textbf{\bibinfo{volume}{404}},
  \bibinfo{pages}{1879} (\bibinfo{year}{2010}), \eprint{1001.4058}.

\bibitem[{\citenamefont{Jeltema and Profumo}(2014)}]{tjspreply}
\bibinfo{author}{\bibfnamefont{T.}~\bibnamefont{Jeltema}} \bibnamefont{and}
  \bibinfo{author}{\bibfnamefont{S.}~\bibnamefont{Profumo}}
  (\bibinfo{year}{2014}), \eprint{1411.1759}.

\bibitem[{\citenamefont{{Jeltema} and {Profumo}}(2014)}]{reply}
\bibinfo{author}{\bibfnamefont{T.}~\bibnamefont{{Jeltema}}} \bibnamefont{and}
  \bibinfo{author}{\bibfnamefont{S.}~\bibnamefont{{Profumo}}},
  \bibinfo{journal}{ArXiv e-prints}  (\bibinfo{year}{2014}),
  \eprint{1411.1759}.

\bibitem[{\citenamefont{{Tamura} et~al.}(2014)\citenamefont{{Tamura}, {Iizuka},
  {Maeda}, {Mitsuda}, and {Yamasaki}}}]{2014arXiv1412.1869T}
\bibinfo{author}{\bibfnamefont{T.}~\bibnamefont{{Tamura}}},
  \bibinfo{author}{\bibfnamefont{R.}~\bibnamefont{{Iizuka}}},
  \bibinfo{author}{\bibfnamefont{Y.}~\bibnamefont{{Maeda}}},
  \bibinfo{author}{\bibfnamefont{K.}~\bibnamefont{{Mitsuda}}},
  \bibnamefont{and} \bibinfo{author}{\bibfnamefont{N.~Y.}
  \bibnamefont{{Yamasaki}}}, \bibinfo{journal}{ArXiv e-prints}
  (\bibinfo{year}{2014}), \eprint{1412.1869}.

\bibitem[{\citenamefont{{Hayato} et~al.}(2010)\citenamefont{{Hayato},
  {Yamaguchi}, {Tamagawa}, {Katsuda}, {Hwang}, {Hughes}, {Ozawa}, {Bamba},
  {Kinugasa}, {Terada} et~al.}}]{2010ApJ...725..894H}
\bibinfo{author}{\bibfnamefont{A.}~\bibnamefont{{Hayato}}},
  \bibinfo{author}{\bibfnamefont{H.}~\bibnamefont{{Yamaguchi}}},
  \bibinfo{author}{\bibfnamefont{T.}~\bibnamefont{{Tamagawa}}},
  \bibinfo{author}{\bibfnamefont{S.}~\bibnamefont{{Katsuda}}},
  \bibinfo{author}{\bibfnamefont{U.}~\bibnamefont{{Hwang}}},
  \bibinfo{author}{\bibfnamefont{J.~P.} \bibnamefont{{Hughes}}},
  \bibinfo{author}{\bibfnamefont{M.}~\bibnamefont{{Ozawa}}},
  \bibinfo{author}{\bibfnamefont{A.}~\bibnamefont{{Bamba}}},
  \bibinfo{author}{\bibfnamefont{K.}~\bibnamefont{{Kinugasa}}},
  \bibinfo{author}{\bibfnamefont{Y.}~\bibnamefont{{Terada}}},
  \bibnamefont{et~al.}, \bibinfo{journal}{Astrophys.J.}
  \textbf{\bibinfo{volume}{725}}, \bibinfo{pages}{894} (\bibinfo{year}{2010}),
  \eprint{1009.6031}.

\bibitem[{\citenamefont{{Nomoto} et~al.}(1984)\citenamefont{{Nomoto},
  {Thielemann}, and {Yokoi}}}]{1984ApJ...286..644N}
\bibinfo{author}{\bibfnamefont{K.}~\bibnamefont{{Nomoto}}},
  \bibinfo{author}{\bibfnamefont{F.-K.} \bibnamefont{{Thielemann}}},
  \bibnamefont{and} \bibinfo{author}{\bibfnamefont{K.}~\bibnamefont{{Yokoi}}},
  \bibinfo{journal}{Astrophys.J.} \textbf{\bibinfo{volume}{286}},
  \bibinfo{pages}{644} (\bibinfo{year}{1984}).

\bibitem[{\citenamefont{{Iwamoto} et~al.}(1999)\citenamefont{{Iwamoto},
  {Brachwitz}, {Nomoto}, {Kishimoto}, {Umeda}, {Hix}, and
  {Thielemann}}}]{1999ApJS..125..439I}
\bibinfo{author}{\bibfnamefont{K.}~\bibnamefont{{Iwamoto}}},
  \bibinfo{author}{\bibfnamefont{F.}~\bibnamefont{{Brachwitz}}},
  \bibinfo{author}{\bibfnamefont{K.}~\bibnamefont{{Nomoto}}},
  \bibinfo{author}{\bibfnamefont{N.}~\bibnamefont{{Kishimoto}}},
  \bibinfo{author}{\bibfnamefont{H.}~\bibnamefont{{Umeda}}},
  \bibinfo{author}{\bibfnamefont{W.~R.} \bibnamefont{{Hix}}}, \bibnamefont{and}
  \bibinfo{author}{\bibfnamefont{F.-K.} \bibnamefont{{Thielemann}}},
  \bibinfo{journal}{ApJS} \textbf{\bibinfo{volume}{125}}, \bibinfo{pages}{439}
  (\bibinfo{year}{1999}), \eprint{astro-ph/0002337}.

\bibitem[{\citenamefont{{Seitenzahl} et~al.}(2013)\citenamefont{{Seitenzahl},
  {Ciaraldi-Schoolmann}, {R{\"o}pke}, {Fink}, {Hillebrandt}, {Kromer},
  {Pakmor}, {Ruiter}, {Sim}, and {Taubenberger}}}]{2013MNRAS.429.1156S}
\bibinfo{author}{\bibfnamefont{I.~R.} \bibnamefont{{Seitenzahl}}},
  \bibinfo{author}{\bibfnamefont{F.}~\bibnamefont{{Ciaraldi-Schoolmann}}},
  \bibinfo{author}{\bibfnamefont{F.~K.} \bibnamefont{{R{\"o}pke}}},
  \bibinfo{author}{\bibfnamefont{M.}~\bibnamefont{{Fink}}},
  \bibinfo{author}{\bibfnamefont{W.}~\bibnamefont{{Hillebrandt}}},
  \bibinfo{author}{\bibfnamefont{M.}~\bibnamefont{{Kromer}}},
  \bibinfo{author}{\bibfnamefont{R.}~\bibnamefont{{Pakmor}}},
  \bibinfo{author}{\bibfnamefont{A.~J.} \bibnamefont{{Ruiter}}},
  \bibinfo{author}{\bibfnamefont{S.~A.} \bibnamefont{{Sim}}}, \bibnamefont{and}
  \bibinfo{author}{\bibfnamefont{S.}~\bibnamefont{{Taubenberger}}},
  \bibinfo{journal}{MNRAS} \textbf{\bibinfo{volume}{429}},
  \bibinfo{pages}{1156} (\bibinfo{year}{2013}), \eprint{1211.3015}.

\bibitem[{\citenamefont{{Travaglio} et~al.}(2004)\citenamefont{{Travaglio},
  {Hillebrandt}, {Reinecke}, and {Thielemann}}}]{2004A&A...425.1029T}
\bibinfo{author}{\bibfnamefont{C.}~\bibnamefont{{Travaglio}}},
  \bibinfo{author}{\bibfnamefont{W.}~\bibnamefont{{Hillebrandt}}},
  \bibinfo{author}{\bibfnamefont{M.}~\bibnamefont{{Reinecke}}},
  \bibnamefont{and} \bibinfo{author}{\bibfnamefont{F.-K.}
  \bibnamefont{{Thielemann}}}, \bibinfo{journal}{A\&A}
  \textbf{\bibinfo{volume}{425}}, \bibinfo{pages}{1029} (\bibinfo{year}{2004}),
  \eprint{astro-ph/0406281}.

\bibitem[{\citenamefont{{Maeda} et~al.}(2010)\citenamefont{{Maeda},
  {R{\"o}pke}, {Fink}, {Hillebrandt}, {Travaglio}, and
  {Thielemann}}}]{2010ApJ...712..624M}
\bibinfo{author}{\bibfnamefont{K.}~\bibnamefont{{Maeda}}},
  \bibinfo{author}{\bibfnamefont{F.~K.} \bibnamefont{{R{\"o}pke}}},
  \bibinfo{author}{\bibfnamefont{M.}~\bibnamefont{{Fink}}},
  \bibinfo{author}{\bibfnamefont{W.}~\bibnamefont{{Hillebrandt}}},
  \bibinfo{author}{\bibfnamefont{C.}~\bibnamefont{{Travaglio}}},
  \bibnamefont{and} \bibinfo{author}{\bibfnamefont{F.-K.}
  \bibnamefont{{Thielemann}}}, \bibinfo{journal}{Astrophys.J.}
  \textbf{\bibinfo{volume}{712}}, \bibinfo{pages}{624} (\bibinfo{year}{2010}),
  \eprint{1002.2153}.

\end{thebibliography}
\end{document}